\documentclass{aa}  

\usepackage{graphicx}
\usepackage{cuted}
\usepackage{txfonts}
\usepackage{hyperref}
\begin{document}

   \title{The model fitting of Gaia DR3 Classical Cepheid light and radial velocity curves.}

   \subtitle{}

   \author{R. Molinaro
          \inst{1}
          M. Marconi\inst{1}
          G. De Somma
          \inst{1,2,3,5}
          V. Ripepi
          \inst{1}
          S. Leccia
          \inst{1}
          I. Musella
          \inst{1}
          T. Sicignano
          \inst{4,5,1,3}
          E. Trentin
          \inst{1}
        M. Gatto
          \inst{1}
          }
   \institute{INAF-OACN Osservatorio Astronomico di Capodimonte, Salita Moiariello 16, 80131, Napoli (ITALY)\\
              \email{roberto.molinaro@inaf.it}
         \and
             INAF-Osservatorio Astronomico d'Abruzzo Via Maggini sn 64100 Teramo, Italy
        \and
Istituto Nazionale di Fisica Nucleare, Sezione di Napoli, Complesso Universitario di Monte S. Angelo, Via Cinthia Edificio 6 I-80126 Napoli, Italy
                      \and
European Southern Observatory, Karl-Schwarzschild-Strasse 2, 85748 Garching bei München, Germany
                 \and
Scuola Superiore Meridionale, Largo San Marcellino10 I-80138 Napoli, Italy
\and
Leibniz-Institut f\"{u}r Astrophysik Potsdam (AIP), An der Sternwarte 16, D-14482 Potsdam, Germany
              \and
Institut für Physik und Astronomie, Universität Potsdam, Haus 28, Karl-Liebknecht-Str. 24/25, D-14476 Golm (Potsdam), Germany
             }

   \date{Received September 15, 1996; accepted March 16, 1997}

% \abstract{}{}{}{}{} 
% 5 {} token are mandatory
 
  \abstract
  % context heading (optional)
  % {} leave it empty if necessary  
   {Classical Cepheids are fundamental astrophysical laboratories for studying stellar structure and evolution, as well as for calibrating the cosmic distance scale. Despite significant progress in observational and theoretical studies, uncertainties remain regarding their masses, luminosities, and distances, as well as the role of processes like core overshooting, rotation, and mass-loss. The advent of high-precision data from ESA Gaia's third data release provides an opportunity to address these questions.}
  % aims heading (mandatory)
   {The primary aim of this study is to estimate the main structural parameters and distances of a sample of Classical Cepheids using non-linear convective pulsational models. The work also seeks to test the consistency of Gaia parallaxes, independently constrain the Mass-Luminosity (M-L) relation, and investigate the dependence of the projection factor (p-factor) on the pulsational period.}
  % methods heading (mandatory)
   {A sample of 46 Classical Cepheids with precise photometric and radial velocity data from Gaia DR3 was analysed. Model fitting was conducted by directly comparing predicted and observed variations in Gaia $\rm G$, $\rm G_{BP}$, and $\rm G_{RP}$ light curves, as well as radial velocity time series. Distances inferred from the models were compared to Gaia parallaxes, including corrections provided by the Gaia team. Predicted masses and luminosities were used to constrain the M-L relation, while the inclusion of radial velocity curves allowed for an independent estimation of the p-factor.}
  % results heading (mandatory)
   {The comparison between inferred distances and Gaia parallaxes reveals statistical agreement, indicating no need for additional global offset corrections. The predicted masses and luminosities are consistent with an evolutionary scenario that includes a small or mild amount of core overshooting, mass loss, or rotation. Our analysis of the p-factor does not suggest a significant period dependence, with a constant value of $p=1.22 \pm 0.05$, consistent with recent literature. Additionally, our results align well with the recent Period-Wesenheit-Metallicity relation derived from Gaia DR3 photometric magnitudes combined with parallax measurements.}
  % conclusions heading (optional), leave it empty if necessary 
   {}

               \titlerunning{The model fitting of Gaia DR3 Classical Cepheid}
               \authorrunning{Molinaro et al.}
   \maketitle

%
%-------------------------------------------------------------------
\section{Introduction}

Classical Cepheids (CCs) are the most important primary distance indicators in calibrating the cosmic distance scale. The physical basis of the existence of Period-Luminosity (PL) and Period-Luminosity-Colour (PLC) relations for these stars is well known. The existence of a Period-Luminosity-Temperature-Mass (PTLM) relation directly descends from the Period-Mean density relation and the Stefan-Boltzmann law \citep[see e.g.][and references therein]{Bono1999_ApJSS_122_167,  Marconi2013_ApJL_768_L6}. Moreover, it is well known from stellar evolution models, that intermediate-mass stars obey a Mass-Luminosity (ML) relation, depending on the chemical composition,  during their central He-burning phase \citep[see e.g.][]{Chiosi1992_ApJ387_320, Chiosi1993_ApJ_86_541, Bono1999_ApJSS_122_167}.  According to several authors \citep[see e.g.][and references therein]{Chiosi1993_ApJ_86_541, Bono1999_ApJSS_122_167, Anderson2014_AA_564_100, Anderson2016_AA_591_8}, the presence of non-canonical effects (e.g. core overshooting, mass loss, and rotation) results in a systematic increase of the luminosity obtained from the "canonical" ML relation. For example, in the recent works by \citet{DeSomma2020JSS_247_30, DeSomma2022_ApJSS_262_25}, the authors considered two over-luminosity levels above the canonical relation (+0.2 dex and +0.4 dex) to take into account non-canonical effects. 
By combining the PTLM with the ML relation, one obtains a Period-Luminosity-Temperature (PLT) or Period-Luminosity-colour (PLC) relation. The widely adopted PL relations result from the projection of the PLC relation onto a PL plane \citep[see][]{Madore1991_PASP_103_933}, so that it implies neglecting the finite colour extension of the instability strip.
Another relation, the so-called Period-Wesenheit relation \citep[PW][]{Madore1982_ApJ_253_575}, is obtained by replacing the magnitude with a function defined as the magnitude corrected for a colour term, multiplied by a coefficient that is exactly the ratio between the total and the selective extinction. This function is reddening-free by definition, provided that an extinction law is assumed a priori.
Thus the coefficients of the ML relation affect the PL, PLC, and PW relations and, in turn, the Cepheid distance scale.
In this context, the combination of accurate geometric distance information and independent pulsation modelling results is needed to constrain the ML relation.
The third ESA Gaia Data Release (Gaia DR3) includes a sample of 799 Cepheids and 1096 RR Lyrae \citep[see e.g.][and references therein]{Katz2023SpecialVelocities, Clementini2023_AA_674_A18, Ripepi2023A&A...674A..17R}, with parallaxes, epoch photometry and radial velocity ($\rm RV$).
The adopted dataset is the fundamental basis to investigate  Cepheid stellar parameters.

From the theoretical point of view, the use of non-linear convective pulsation models 
to compare predicted and observed light and RV curves, represent a powerful tool to simultaneously constrain the intrinsic stellar parameters and the individual distances of the investigated pulsating stars.
Several applications in the literature supported the predictive capabilities of this method both for the CCs \citep[see e.g.][]{Wood1997ApJ...485L..25W, Bono2002ApJ...565L..83B,Keller_2002_Apj_578_144,Keller2006, Natale2008_ApJ_674_L93,Marconi2013_ApJL_768_L6,Marconi2013_MNRAS_428_2185,Marconi2017_mnras_466_3206,Ragosta2019_mnras_490_4975} and RR Lyrae \citep[see e.g.][]{Bono2000_ApJ_532_L129,Marconi2005_AJ_129_2257,Marconi2007_AA_474_557}.

These pulsation codes are not only capable of prediction of stellar parameters, but they also have been able to describe dynamical phenomena of these stars: e.g. beat Cepheids \citep{kollath_2002A&A...385..932K}, double mode RR Lyrae behaviour \citep{szabo_2004A&A...425..627S}, period doubling and related non-linear phenomena \citep{kollath_2011MNRAS.414.1111K, plachy_2013AN....334..984P, smolec_2014IAUS..301..265S, smolec_2016CoKon.105...69S}. Meanwhile, PL relations are extensively studied as well \citep[see e.g.][]{Marconi2015_ApJ_808_50, marconi_2018ApJ...864L..13M, deka_2022MNRAS.517.2251D, das_2021MNRAS.501..875D, das_2024A&A...684A.170D}

In this work, we used a new extensive set of non-linear convective pulsation models to fit a sub-sample of 46 Fundamental CCs from the Gaia DR3 dataset. This represents the largest sample of CCs analysed so far through the model fitting technique.

The work is structured as follows: in Section \ref{sec:sample-selection} we present the adopted sample, whereas the new computed pulsation models are discussed in Section \ref{sec:pulsational-models}; the developed model-fitting approach is outlined in Section \ref{sec:model fitting method}, together with the selection of the best-fit models based on the inclusion of Monte Carlo simulations; Section \ref{sec:results of the fit} presents the results of the fitting method; lastly Section \ref{sec:conclusions} is dedicated to the closing remarks. 

\section{The sample selection}\label{sec:sample-selection}

The initially adopted sample of sources was obtained from a query to the Gaia public archive\footnote{\href{https://gea.esac.esa.int/archive/}{https://gea.esac.esa.int/archive/}}.  All the Data Release 3 (DR3) fundamental mode CCs, also having epoch RV data, were extracted and amounted to  799 variables. This catalogue was cross-matched with the list of confirmed Galactic CCs by \citet{Ripepi2023A&A...674A..17R}. This cross-match left us with 558 variables. Since the pulsational model calculation is very time-consuming, we decided to reduce the sample by selecting only those having a good phase sampling in the $RV$ curve\footnote{We note here that after 34-month operations, the $G$, $G_{BP}$ and $G_{RP}$ curves are almost all well covered in term of phase.}. To this aim, we modelled the $RV$ time series by fitting them with a truncated Fourier series (App.~\ref{app:initial-fourier-model}) and selected only those without large/spurious model oscillations, possibly due to wide phase intervals with a dearth of measures.  
In particular, a selection was performed  considering only sources according to the following conditions: i) RVS curves with at least 20 data points; ii) an uniformity index\footnote{A measure of the non-redundancy of the phase coverage and of the uniformity of the realized phase sampling. It assumes values between 0 and 1, \citep{Madore2005_ApJ_630_1054}.} $UI_{RV}>0.9$; iii) visual inspection to remove the remaining few bad sources; iv) DR3 relative parallax error $\sigma_\varpi/\varpi<50\%$.

After this procedure, the sample was reduced to 198 selected sources. 
This sample was further reduced by looking at the residuals around the Fourier models, for both photometric and RV data. Specifically, we performed a statistical analysis of the Fourier fit scatter for all time series. In particular, we calculated the root mean square (rms) of residuals around the Fourier model for all the considered photometric and RV curves and removed all the sources with the rms of residuals deviating by more than 3.5 Median Absolute Deviation $(MAD)$ from the median of the rms distribution, for one or more observed time series. The remaining sample after this step contains 46 sources.

We then complemented the information with available metal abundance measurements, as metallicity is a key parameter affecting stellar pulsation and the shape of the light and radial velocity variations. We adopted the work by \citet[][T24]{Trentin2024_AA_681_A65} which reports the iron abundance for more than 900 CCs obtained based on high-resolution spectroscopy. 

\begin{figure}
    \centering
    \includegraphics[height=8cm]{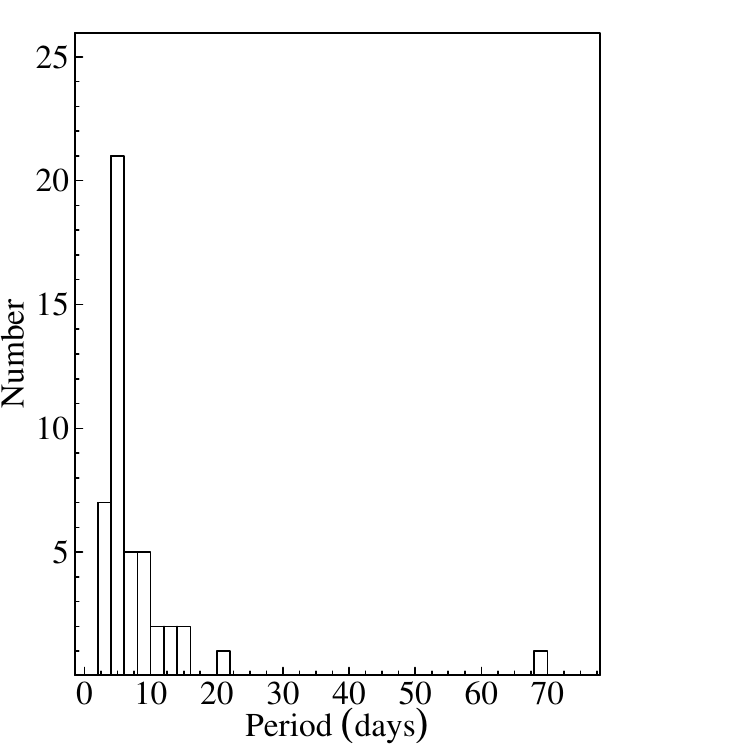}
    \caption{The distribution of the selected DCEP period values obtained from the DR3.}
    \label{fig:finalSamplePeriod}
\end{figure}

\begin{figure}
    \centering
    \includegraphics[height=8cm]{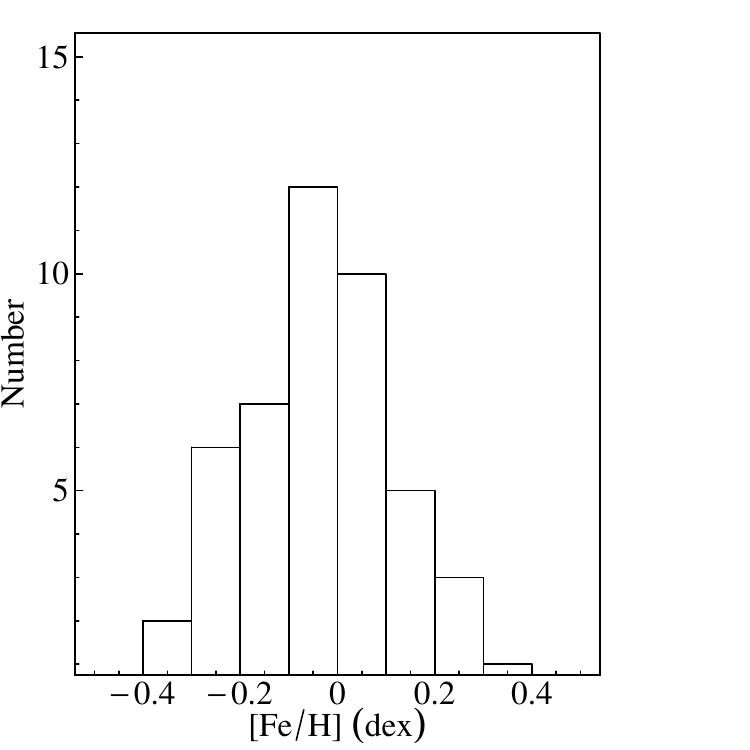}
    \caption{The distribution of the selected DCEP $[Fe/H]$ values, obtained through high resolution spectroscopy (T24).}
    \label{fig:finalSampleFeH}
\end{figure}

The period and the $[Fe/H]$ distributions of our final sample are shown respectively in fig.~\ref{fig:finalSamplePeriod} and fig.~\ref{fig:finalSampleFeH}.

\begin{figure}
    \centering
    \includegraphics[height=8cm]{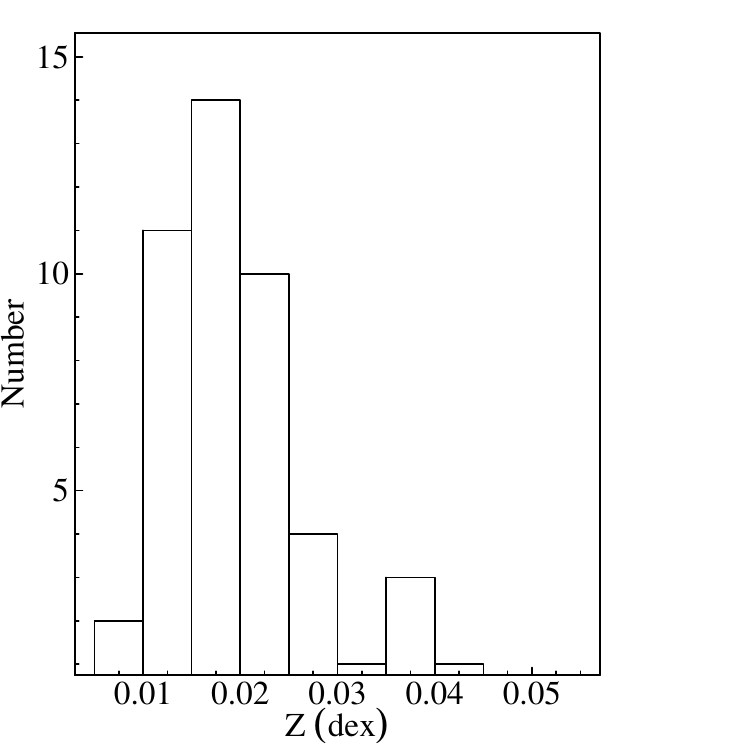}
    \caption{The distribution of the Z values of all the selected variables, computed by transforming the spectroscopic $[Fe/H]$ values from \citet{Trentin2024_AA_681_A65}.}
    \label{fig:finalSampleZ}
\end{figure}

Assuming the elemental composition of the Sun $Z_\odot = 0.02$ dex \citep[see e.g.][and references therein]{Buldgen_2024_AA681_A57}, we converted the observational $[Fe/H]$ values into Z\footnote{$\rm Z = [Fe/H] + \log_{10}(Z_\odot)$}, obtaining the distribution shown in Fig.~\ref{fig:finalSampleZ}.

\section{Pulsational models computations} \label{sec:pulsational-models}
The theoretical framework considered in this work is based on a non-linear radial pulsation code that adopts a non-local time-dependent treatment of the turbulent convection \citep[see e.g.][and references therein]{Stellingwerf1982, Bono1994a, Bono1999_ApJSS_122_167}. 

The procedure followed to obtain the hydrodynamic pulsational models reproducing the observed light and radial velocity curves, consists of four main steps: i) the definition of a multi-dimensional grid of possible structural parameter ($\rm M$, $\rm T_{eff}$, $\rm \log(L/L_\odot)$, $\rm \alpha$\footnote{It is defined by the ratio between the mixing length and the pressure height scale}) values to be given as input to the hydrodynamical code to 
reproduce the observed period; ii) the linear analysis that provides the static envelope structure to be perturbed through the subsequent non-linear analysis; iii) the non-linear analysis aiming at producing the final full-amplitude pulsating models, including both absolute bolometric magnitude and pulsational velocity curves iv) the transformation of the theoretical bolometric light curves into the Gaia photometric system.

\subsection{Initial grid of structural parameters}\label{sec:initial-grid}
Since chemical composition plays a key role in the computation of pulsation models, and given that the selected sample spans a wide range in Z, from $\rm Z<0.01$ dex to $\rm Z\simeq 0.045$, we decided to divide the sample into four metallicity bins centred around the values $\rm Z_1=0.01$ dex, $\rm Z_2=0.02$ dex, $\rm Z_3=0.03$ dex, and $\rm Z_4=0.04$ dex, each characterized by a width of 0.005 dex. The Helium content Y, corresponding to the adopted Z bins, are respectively equal to $\rm Y_1=0.26$, $\rm Y_2=0.28$, $\rm Y_3=0.28$ and $\rm Y_4=0.29$ dex. Based on this division, each bin contains 13, 24, 5, and 4 sources, respectively. From this point forward, the central value of each bin will be representative of the elemental composition of the stars within that bin. 

For each fixed chemical composition, the input mass and effective temperature values for the linear analysis were varied on a rectangular grid, covering the ranges 3$\rm <M/M_\odot<$13 with a step of 0.5 $M_\odot$ and 3500$\rm <T_{eff}(K)<$7500 with a step of 50 K, respectively. Then, for each combination of mass and effective temperature, the luminosity value is found by satisfying the observed period, as described in the Appendix \ref{app:ML-PTLM equations}. In this procedure only luminosities satisfying a variation  within $-0.06$ dex to $0.6$ dex with respect to the canonical ML relation, as predicted by \citet{Bono_2000_ApJ_543_955}, are kept for the following analysis (see Appendix \ref{app:ML-PTLM equations}). The lower limit corresponds to -3 times the standard deviation of the adopted ML relation \citep{Bono_2000_ApJ_543_955}, while the upper limit takes into account the presence of full core overshoot \citep{Chiosi1993_ApJ_86_541}. Taking advantage of the well known relations between luminosity and the other structural parameters, our method allows us to define a complete grid of initial structural parameters, by significantly reducing  the number of models to be computed.

As regards to  the $\alpha$ parameter, it was varied between 1.4 and 1.9 in steps of 0.05, resulting in 11 possible $\alpha$ values. We note that when the $\alpha$ values do not match those provided in Table 3 of \citet{DeSomma2020JSS_247_30}, we applied linear interpolation if our values fell between two of those considered by the authors, or we adopted their most extreme relations if our values lay outside their specified $\alpha$ range.

\subsection{The hydrodynamical analysis}
The linear survey is based on a non-adiabatic radiative code \citep[see][and references therein, for detail]{Bono1999_ApJSS_122_167} that was run for all the M-$\rm T_{eff}$-$\rm \log(\rm L/L_\odot)$ values defined above for each assumed chemical composition. 

The subsequent non-linear analysis takes the static structure and the predicted eigenfunctions obtained from the linear code as input and enables the examination of which models exhibit a full amplitude stable pulsational behaviour.

The non-linear code introduces an initial dynamical perturbation (typically 20 km/sec) to the static model, modulated within the envelope structure with the predicted radial eigenfunction obtained from the linear analysis.  Subsequently, it monitors the time evolution of its characteristic parameters to assess the attainment of full amplitude behaviour. 

Integration of hydrodynamical equations is performed over a considerable number of periods, with the satisfaction of convergence conditions determined by the total work approaching vanishing values and the peak-to-peak bolometric amplitude exhibiting negligible change (see below).

Typically, the code is initiated with a fixed number of computing cycles, ranging from 1000 to around 10000, depending on the model's mass and position within the instability strip, with models located close to the extreme edges of the strip typically requiring a larger number of cycles. Given the resource-intensive nature of the non-linear step, we have implemented runtime tests within the code to assess the model's convergence status. Specifically, the non-linear code computation is halted under the following circumstances over the last 40 period cycles:

i) The normalized total work remains stable around vanishing values ($\leq5\cdot 10^{-4}$), and the peak-to-peak bolometric magnitude amplitude shows minimal variation, staying below $5\cdot 10^{-3}$ mag.

ii) The peak-to-peak amplitude is consistently smaller than 0.02 mag and steadily decreasing.

iii) The conditions described in point (i) are met, but the period converges to the expected value associated with another pulsation mode (the model reaches a stable pulsational regime but with a mode that is not the required one).

Upon meeting the conditions in point i), the model achieves full amplitude behaviour, pulsating stably. Conversely, if conditions ii) or iii) are satisfied, the pulsation ceases or the model switches to a different pulsational mode.

This strategy, allows us to speed up the model computation by stopping it before the assumed total number of cycles is reached, as soon as the convergence condition is achieved, and in particular when the pulsation behaviour switches off or changes mode.

\subsection{The transformations from bolometric to Gaia multi-filter light curves}

The conversion of the bolometric magnitude curves into the Gaia photometric bands was achieved by adopting the bolometric correction (BC) grids based on the PHOENIX model atmosphere \citep{Chen2019_AA_632_A105}. We refer the reader to the work by \citet{Marconi2021_mnras_500_5009} for all the details of our adopted procedure. Here we just recall that we used a proprietary {\tt C} code to interpolate the \citet{Chen2019_AA_632_A105} tables to obtain the BCs corresponding to the $\rm \log(g)$, $\rm T_{eff}$, and Z of our modelled curves. For all the transformed models we assumed $\rm M^{bol}_\odot = 4.79$ mag \citep[][and references therein]{DeSomma2022_ApJSS_262_25}.

\section{The model fitting method}
\label{sec:model fitting method}

The adopted model fitting procedure is based on the minimization of the following $\chi^2$ functions \citep[see also][]{Marconi2013_ApJL_768_L6,  Marconi2017_mnras_466_3206, Ragosta2019_mnras_490_4975}:
\begin{eqnarray}
\rm \chi_{j=1\dots N_{bands}}^2 = \sum^{N^j_{pts}}_{i=1}\left(\frac{m_i^j\left(\phi_i^j\right) - M^j\left(\phi_i^j + \delta \Phi^j\right) + \mu^j}{\sigma^j_i}\right)^2 + \chi^2_{Amp^j} \label{eq:chi2Phot} \\
\chi_{RV}^2 = \sum^{M_{pts}}_{l=1}\left(\frac{v_l\left(\phi_l\right) - a\left(V_{puls}\left(\phi_l + \delta \Phi_{RV}\right) + \gamma \right)}{\sigma^{RV}_l}\right)^2 + \chi^2_{Amp_{RV}} \label{eq:chi2RV}
\end{eqnarray}
where we included the term $\chi^2_{Amp}$ for every fitted band (including RV), which is equal to:
\begin{equation}\label{eq:chi2-amp-termp}
   \rm \chi^2_{Amp^j} =\left ( \frac{Amp^j_{Model} - Amp^j_{Fourier}}{\sigma_{Amp}^j}\right)^2
\end{equation}
To justify the inclusion of this additional term, we performed tests which revealed that, in some cases, models minimizing the $\chi^2$ function without the $\rm \chi^2_{Amp}$ term, do not always reproduce the peak-to-peak amplitude of the observational curves. We therefore decided to give greater weight to models that also reproduce as good as possible the amplitude, whose reference value ($Amp^j_{Fourier}$) is derived from the Fourier model constructed as previously described.

The first equation above refers to the j-th photometric band, while the
second equation is used to fit the radial velocity curve. The sums are performed over the number of observational data points for both photometry ($N_{pts}$) and radial velocity ($M_{pts}$). 
The lower case symbols, $\phi$, $m$  and $v$ indicate the pulsation phases, the measured magnitude and the measured radial velocity, respectively. The  upper case  functions $M^j(\phi)$ and $V_{puls}(\phi)$
give analytic expressions for the model j-th photometric
band and the pulsational velocity curves\footnote{A smoothing--spline is used to provide analytical expressions of  the  light and  pulsational velocity
curves of the fitted models.}. The expected outputs of this best-fitting procedure are the distance modulus, $\rm \mu^j$, and the phase shift
$\rm \delta \Phi^j$  in the j-th band, for the photometry, while the model fitting of the radial velocity curves provides information on the projection factor $\rm \left (p = -\frac{1}{a}\right )$,
the  velocity of  the  centre of mass $\gamma$ and the phase shift
$\rm \delta \Phi_{RV}$. 

The projection factor p allows us to compare the modelled pulsational velocity to spectroscopically measured radial velocity. It quantifies the projection of the pulsational velocity of the stellar atmosphere along the line of sight. In particular it is defined as $p=V_{puls}/RV$ and assumes values larger than 1.0. The value of the $p$ factor is debated in the literature due to the difficulty of its determination. Indeed, it depends on the physical structure of the stellar atmosphere, on the velocity gradient in the atmosphere and on the way the radial velocity is obtained experimentally \citep[see e.g.][and references therein]{Nardetto2011TheCepheids, Molinaro_2011_mnras_413_942, Ragosta2019_mnras_490_4975}. 
The first works addressing the problem of the projection factor from a theoretical perspective date back to Eddington (1926) and Carroll (1928). These authors, by including the effects of limb darkening and constant-velocity expansion of the stellar atmosphere in their theoretical models, obtained a projection factor value of p=24/17=1.41. This value is close to the maximum limit, 1.5, which is derived when the Cepheid is approximated as a simple uniform disk \citep{Nardetto_2017_Proc_IAU_Symp_330}. A more complex model, which accounts not only for the geometric effect but also for the relationship between the stellar surface's radial velocity and that derived from spectral lines, was developed for example by \citet{Nardetto_2004_AA_131_137}. This model also includes the effects of limb darkening and atmospheric velocity gradients. Several studies have investigated the existence of a relationship between the p-factor and the pulsation period, but no definitive conclusion has been reached so far \citep{Groenewegen_2007_AA_474_975, Groenewegen_2013_AA_550_70, Kervella_2017_AA_600_127, Nardetto_2009_AA_502_951, Neilson_2012_AA_541_134, Storm2011}. 

In the current work, we forced the p-factor values in the range of 1.10 -- 1.5, to take into account current uncertainties in the literature evaluations. The upper limit was fixed according to the expected maximum theoretical value, as stated above, while the lower limit was adopted according to the recent results obtained by \citet{Trahin_2021_AA_656_102}. They calibrated the p-factor value by combining the parallax-of-pulsation method, implemented through the SPIPS algorithm \citep{Merand2015_AA_584_80}, with the Gaia EDR3 parallaxes  \citep[see also][]{Anderson2024VELOcitiesCepheids, Breitfelder_2016_AA_587_117, Trahin_2021_AA_656_102}.

The fitted phase shifts are introduced to take
into account small uncertainties in the estimate of the phases of the
observations, and are typically around $\sim 0.02$.

In the current work, the $\chi^2$ minimum value was obtained using an analytical approach, rather than relying on commonly used numerical minimization routines \citep[e.g.][]{Levenberg_1944}. Specifically, we show that minimizing the $\chi^2$ function in eq.\ref{eq:chi2Phot} and eq.\ref{eq:chi2RV} can be reduced to solving two equations for the unknowns $\delta \Phi^j$ and $\delta \Phi_{RV}$, while the remaining fit parameters can be expressed as functions of these phase shifts (see App.~\ref{app-chi2-minimization}). This approach was feasible in our specific case because we were able to derive the first-order analytical derivatives and solve the resulting system of equations explicitly.

The minimization procedure is applied separately to each photometric band and to radial velocity, and finally, the obtained minimum $\chi^2$ values are combined into one cumulative minimum value according to the following relation:
\begin{equation}
\rm    \chi^2_{comb} = 1/N_{bands}\sum_{j=1}^{N_{bands}}\chi^2_j/(d.o.f.)_j\label{eq:combinedChi2}
\end{equation}
where $N_{bands}$ is the number of available bands (including RV), 
while $\rm (d.o.f.)_j$ represents the number of degrees of freedom for the $j$-th band\footnote{In the case of photometric bands we have $\rm \nu_j=N^j_{pts}-3$, while for RV $\nu_{RV}=N^{RV}_{pts}-4$, where we took into account the presence of a further degree of freedom for both photometry and radial velocity, due to the $\chi^2_{Amp}$ term.}.

Here we want to focus the attention on the fact that equations \ref{eq:combinedChi2} was minimized at fixed $\alpha$ value, for all the possible values we have used to divide the range 1.4-1.9 (see sec.~\ref{sec:initial-grid}). Finally,the optimal $\alpha$-value was determined as described below.
For each candidate $\rm \alpha_i\, i=1...11$, we:
\begin{itemize}
    \item Selected the corresponding subsample of pulsation models.
    \item Performed independent fits by minimizing the $\rm \chi^2_{comb}$ function (Eq. \ref{eq:combinedChi2}).
    \item Recorded the resulting best-fit $\rm \chi^2_i$ value.
\item  The final $\alpha$ was chosen as the value satisfying $\rm \chi^2_{best} = min{\chi^2_{\alpha_1}, ..., \chi^2_{\alpha_{11}}}$.
\end{itemize}

\section{Monte Carlo Methods for best model selection and uncertainty quantification}\label{sec:MonteCarlo}
Monte Carlo techniques encompass a broad class of numerical methods based on the random resampling of a dataset to derive statistical estimates (e.g., confidence intervals, error estimations). In this work we used two different kinds of simulations: i) bootstrap simulations, obtained by resampling input time series; ii) Monte Carlo simulations with observational noise injection. 
The following subsections detail each method's implementation and their specific roles in our analysis.

\subsection{Bootstrap resampling}
Among Monte-Carlo simulations, bootstrap resampling is a widely used approach. 
Bootstrap methods come in several variants (e.g., parametric, non-parametric, jackknife), and we refer the reader to the review by \citet{Feigelson2012msma.book.....F} for a comprehensive discussion. In this study, we specifically apply the classical non-parametric bootstrap method. This approach involves generating mock datasets by randomly resampling the original dataset with replacement. Each simulated dataset has the same size as the original set of measurements, but individual data points can appear multiple times, while others may be absent in a given resampled dataset \citep{Feigelson2012msma.book.....F}.

In our work, we performed 250 random simulations. This number was chosen to strike a balance between computational time and the need for a sufficiently large number of trials. Each random experiment consists of two main steps:
\begin{itemize}
    \item Resampling: the observational time series of both photometry and radial velocity are resampled with replacement, ensuring that each mock dataset retains the same number of data points as the original one.
   \item   Model fitting: the full fitting procedure is repeated for each mock dataset.
\end{itemize}
   
This approach serves a dual purpose:
\begin{itemize}
\item For a given n-tuple of structural parameters (M, $\rm T_{eff}$, $\rm log L/L_\odot$)\footnote{This point refers to a fixed value of $\alpha$. See Section~\ref{sec:MonteCarlo} for details on how the best $\alpha$ value was selected.}, corresponding to a fixed stellar model, the simulations allow us to quantify the uncertainties in the fitted parameters derived from the solutions of:
Equations \ref{eq:chi-deltaphi-phot}, \ref{eq:chi-deltaphi-rv}, \ref{eq:chi2-mui}, \ref{eq:chi2-gamma}, \ref{eq:chi2-pfact} (namely namely $\rm \mu_i$, $\rm \phi_i$, $\gamma$, $\rm \phi_{rv}$, and p-factor)).
The uncertainty for each fitted parameter is computed as the robust standard deviation ($\rm 1.4826 \cdot MAD$), which rescales the median absolute deviation to match the standard deviation of a normal distribution. These values are obtained from the 250 best-fit solutions generated by the Monte Carlo realizations.
\item For each of the 250 simulations, the optimal pulsation model is identified as the one minimizing the total $\chi^2$ function defined in Eq.~\ref{eq:combinedChi2}. This yields a distribution of optimal models that typically span multiple nodes of the parameter-space grid. Since the location of the minimum $\chi^2$ varies from simulation to simulation, this dispersion provides a direct measure of the parameter-space degeneracies introduced by observational uncertainties.
This behaviour is a fundamental property of $\chi^2$: if the phase coverage of the data is uneven, the best-covered phases will dominate the $\chi^2$ statistic, effectively weighting those regions more heavily during minimization. Our resampling strategy, combined with the inclusion of the peak-to-peak amplitude term (Eq.~\ref{eq:chi2-amp-termp}), helps to mitigate this effect.
\end{itemize}

Focusing on the last itemized point discussed above, we analysed the distribution of minimum $\chi^2$ values (hereafter $\chi^2_{\rm min}$) obtained from our 250 simulations. A statistical characterization of this distribution enabled us to estimate the expected minimum $\chi^2$ value. We adopted the median as a robust central estimator, and the 5th–95th percentile range as a measure of its uncertainty. Prior to this analysis, significant outliers—defined as values beyond $\pm 3.5 \cdot \mathrm{MAD}$—were excluded to ensure statistical reliability.
All models contributing to the clipped $\chi^2_{\rm min}$ distribution were then used to infer the optimal structural parameters and their associated uncertainties. This subset defines the best-fitting solutions for the observed source. For each structural parameter (mass, $\rm T_{\rm eff}$, and $\log (L/L_\odot)$), the best estimate was taken as the median of its distribution, while the uncertainty was quantified via the semi-range (i.e., half the difference between the maximum and minimum values). This conservative approach was adopted to:
(i) provide a robust estimate that reflects the full variability of viable solutions;
(ii) avoid assumptions regarding the shape of the underlying distributions; and
(iii) prevent underestimation of the true uncertainties.

\subsection{Measurement-Error-Driven Monte Carlo Sampling}
In addition to the bootstrap technique introduced in the previous section, this work employs another Monte Carlo-based approach. Specifically, to estimate the uncertainty associated with a given property in our sample of Cepheids—and to perform a statistical analysis that incorporates the measurement errors affecting each input data point (e.g., p-factor, global parallax offset)—we adopted the following methodology.:
i) we conducted 250 Monte Carlo simulations, perturbing each data point within its error bar under the assumption of Gaussian noise statistics; ii) for each realization, we computed the mean of the studied quantity. The resulting distribution of the 250 mean values was then analysed, and we defined the 5th-95th percentile interval as the best estimate of the uncertainty for the considered quantity.

\section{Results of the fit}\label{sec:results of the fit}
The multi-filter Gaia light curves and the velocity curves of the best models found (lines) for all the selected sources, are shown in Fig. D.1 of the \href{https://doi.org/10.5281/zenodo.15722385}{Appendix D}, over-imposed to the observed photometric and radial velocity time series. Table 1 in \href{https://doi.org/10.5281/zenodo.15722385}{Appendix D} lists the best-fit parameters derived from our analysis. For each star, we report the optimal estimate of the pulsation period ($\rm P_{best}$), the corresponding value of the $\alpha$ parameter, and the central values of metallicity and helium content ($\rm Z_{best},\,Y_{best}$) characterizing the best-fitting chemical composition bin. We also provide the best-fit structural parameters, namely mass (M), effective temperature ($\rm T_{eff}$), and luminosity ($\log(L/L_\odot)$), as well as the over-luminosity ($\rm \Delta_{ML}$) with respect to the canonical ML relation. Additionally, the table includes the best-fit projection factor, the $\gamma$-velocity, the theoretical parallax estimate, and the global $\chi^2$ value obtained by minimizing Eq.~\ref{eq:combinedChi2}

While not perfect, the theoretical framework provides a reasonable match to both light and radial velocity curves, demonstrating its ability to predict multi-filter variations of Cepheids with different periods and metallicities, despite some systematic deviations. Here we list the main features and issues around the fit, hinting that the problem of stellar pulsation is still not solved:
\begin{itemize}
    \item    Systematic larger modelled amplitudes in the photometric Gaia bands than in the RV curves (CK Cam, V407 Cas, IN Aur, V432 Ori, OZ Cas, V737 Cen, S Vul):
These discrepancies could potentially be reduced by fine-tuning the $\alpha$ parameter, as increasing $\alpha$ might yield models with smaller photometric amplitudes, thereby improving the fit. However, raising $\alpha$ would also decrease the p-factor (see Eq.~\ref{eq:chi2RV}), which—especially for stars like IN Aur, V432 Ori, OZ Cas, V737 Cen, and S Vul—could push it toward the lower limit of the physically expected range.
The observed differences likely stem from limitations in the current treatment of convection-pulsation coupling, particularly the challenges associated with turbulent convection modeling \citep{Marconi2017_mnras_466_3206, kovacs_2023MNRAS.521.4878K}. 
\item     Minor phase shifts in secondary features (BF Cas, V384 CMa, AP Pup, WX Pup, OGLE-GD-CEP-0339):
these small discrepancies can rise from limitation in the treatment of the convection-pulsation coupling that affects the relative amplitude and position of the primary and secondary maximum.
\item    Severe mismatches (V339 Cen, HZ Per, SS CMa, AD Pup, RZ Vel, S Vul): the point detailed above is more evident in these cases, that are closer to the red edge of the instability strip and in some cases affected by limitations in the light curve sampling (see e.g. S Vul).  
\end{itemize}
All the cases listed are characterized by a global $\chi^2$ value significantly exceeding the expected value of 1, reflecting notable discrepancies between the model and the observations.
While these issues are essential to resolve for a complete understanding of the underlying physics, they have a minor impact on the determination of structural parameters. In the following sections, we therefore focus on presenting the results from the analysis of the best-fit model properties.

\subsection{Predicted versus measured periods}
\begin{figure}
    \centering
    \includegraphics[height=8cm]{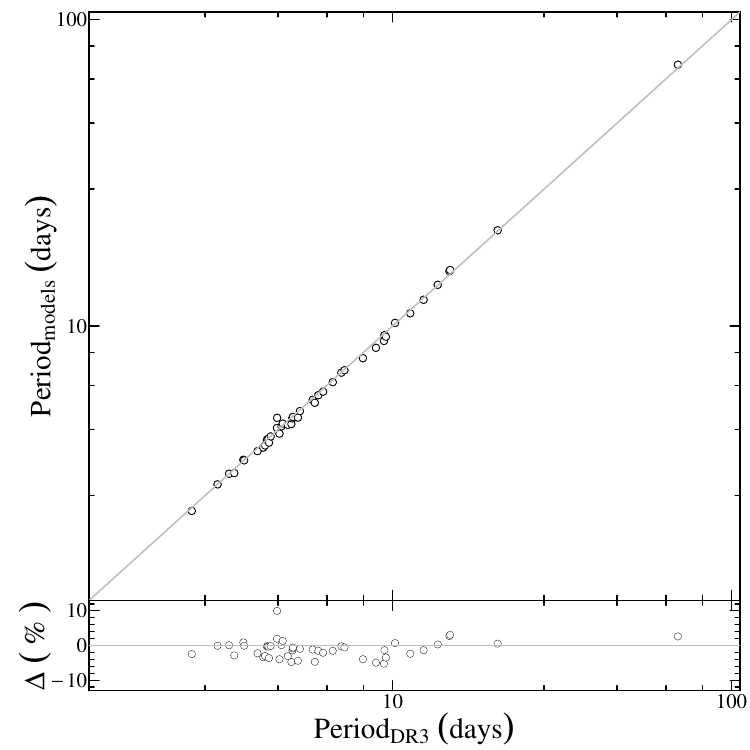}
    \caption{In the top panel, the modelled period values are on the vertical axis, while the DR3 periods on the horizontal one. The 1:1 line is also plotted to facilitate the comparison. The percentage difference between the theoretical periods and the DR3 values is plotted in the bottom panel as a function of the observational period.}
\label{fig:PeriodComparison_fullMwDcepF_Final_singleRun_newSelection}
\end{figure}

In Fig.~\ref{fig:PeriodComparison_fullMwDcepF_Final_singleRun_newSelection}, we compare the pulsational periods based on the best fit models, with the values from Gaia DR3. In the bottom panel of the quoted figure, we show the percentage differences between modelled periods and observational periods. The median and the robust standard deviation  ($\rm 1.4826\cdot MAD$) of the differences are equal to 1.3\% and 1.6\% respectively, while the maximum and minimum differences do not exceed 10\%.

\subsection{The projection factor}\label{sec:p-factor}
As previously noted, the match between the theoretical and observed amplitudes of the radial velocity curves provides the corresponding stellar projection factor.

\begin{figure}
    \centering
    \includegraphics[trim={0cm 6.6cm 0cm 0cm},clip,height=4cm]{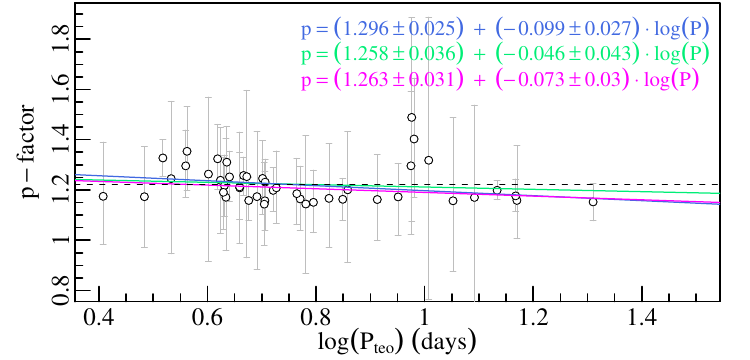}
    \caption{The projection factor as a function of the best fit pulsational period. Three kinds of linear fit are also reported: the weighted linear fit (blue line), the unweighted linear fit (green line), and the robust linear fit (magenta line). The dashed black line indicates the median projection factor computed by considering the whole sample.}
\label{fig:PeriodPfactor_fullMwDcepF_Final_singleRun_newSelection}
\end{figure}

Using the inferred results for the p--factor, we investigate its trend with the period by fitting the equation $\rm p= \alpha + \beta \cdot \log(P)$. The results of this analysis are shown in Fig.~\ref{fig:PeriodPfactor_fullMwDcepF_Final_singleRun_newSelection}.  In the figure, we have highlighted the three different types of fits that were tested. Despite of the fact that the obtained slopes are significantly different from zero, the very small determination coefficient $\rm R^2\sim 0.22$ indicates that the correlation is very weak. Although our analysis does not reveal a statistically significant correlation, large uncertainties associated with several p-factor estimates may hinder the detection of a possible underlying trend.

Neglecting the period dependence, we derived the mean p factor value based on the discussed model fitting results with its associated statistical error: $p=1.22\pm 0.05$. The uncertainty on the projection factor was determined through a set of Monte Carlo simulations. First, we varied the projection factor of each individual source within its error, assuming a Gaussian distribution. For each simulation, we calculated the mean of the resulting projection factors. Finally, the uncertainty was computed as the 5-95 percentile interval of all the mean values obtained.

The lack of a significant dependence of the p-factor on the pulsation period is consistent with the findings of \citet{Nardetto_2009_AA_502_951, Neilson_2012_AA_541_134}, where the p-factor was calibrated using theoretical models.

Similarly, \citet{Groenewegen_2007_AA_474_975} found no significant period dependence, deriving a constant value of p=1.27$\pm$0.05 by applying the Baade-Wesselink method in its surface brightness form. A comparable p-factor of 1.27$\pm$0.06 was also obtained by \citet{Merand2015_AA_584_80} for $\delta$ Cep.

Our estimate of the p-factor is in excellent agreement with the values derived for the binary Cepheid OGLE-LMC-CEP-0227 by \citet{Marconi2013_ApJL_768_L6} (p=1.20$\pm$0.08) and \citet{Pilecki2013} (p=1.21$\pm$0.03$\pm$0.04). This binary Cepheid is particularly significant for studying the structural parameters of Cepheids, as its binary nature provides an independent estimate of the primary characteristics of the variable. In \citet{Marconi2013_ApJL_768_L6}, the variable was modelled using hydrodynamic pulsation models, while \citet{Pilecki2013} introduced a method specifically designed for analysing double-lined eclipsing binaries containing a radially pulsating star.

Our best estimate is close to the lower end of the assumed p-factor range, but we need to extend the sample in order to equally populate the various period ranges in order to provide more stringent results.

\subsection{The $\alpha$ parameter}
The determination of p-factor is strongly affected by the choice of the optimal $\alpha$-parameter. Turbulent convection models based on a single-equation formalism, commonly used in radial pulsation studies, often fail to simultaneously reproduce both the radial velocity and photometric variations when a constant p-factor is assumed, as shown by \citet{kovacs_2023MNRAS.521.4878K}. This limitation also affects our hydrodynamic code and has similarly been observed in Cepheid models by \citet{Marconi_2017EPJWC.15206001M}.

\begin{figure}
    \centering
    \includegraphics[trim={0cm 6.6cm 0cm 0cm},clip,height=4cm]{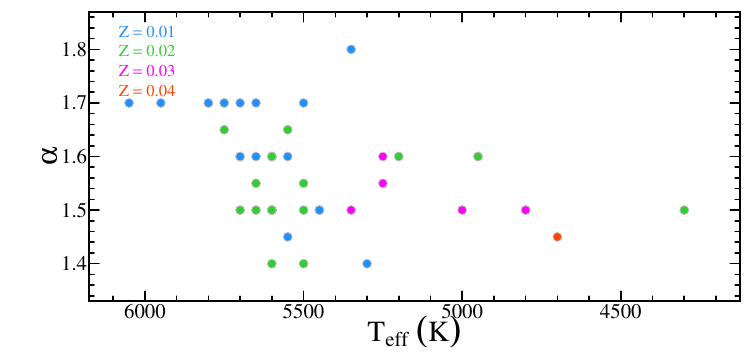}
    \caption{The best-fit $\alpha$ parameter plotted against the best values of the $\rm T_{eff}$. Different colours indicate different metallicity bins.}
\label{fig:alpha_vs_Teff_fullMwDcepF_Final_singleRun_newSelection_Ebv.lit}
\end{figure}

We notice that, on the basis of our results, no clear trend is found for the variation of $\alpha$  with the effective temperature. However, the sample analysed here might be not large enough to investigate such a dependence, and the large scatter visible in the $\alpha$ versus $T_{\rm eff}$ plot (see fig.~\ref{fig:alpha_vs_Teff_fullMwDcepF_Final_singleRun_newSelection_Ebv.lit}) prevents us from drawing firm conclusions. 
Moreover, metallicity affects the predicted amplitudes and in turn the selected  $\alpha$ values, so that differences in metallicity  in some cases contribute to the plot dispersion.
It is nevertheless interesting to note that, previous findings in the literature, noted an anti-correlation between the eddy viscosity and the effective temperature  for RR Lyrae stars by \citet{kovacs_2023MNRAS.521.4878K, Kovacs_2024MNRAS.527L...1K}. Similar results were reported for classical and type-II Cepheids by \citet{deka-2024MNRAS.530.5099D}, although without an in-depth analysis of this aspect.
While one could argue that those previous results were obtained using different codes — namely the Budapest-Florida Code \citep{Yecko_1998A&A...336..553Y} and MESA-RSP \citep{Paxton2019, Smolec_2008AcA....58..233S} — similar trends were also found for RR Lyrae stars and bump Cepheids with the Stellingwerf code \citep{DiCriscienzo_2004ApJ...612.1092D, Bono2002ApJ...565L..83B}.
However all the investigations adopted a single chemical composition and did not take into account simultaneously photometric and radial velocity variations.

\subsection{Absorption}\label{sec:Absorption}
To obtain the reddening-free distance modulus $\mu_0$, we corrected the distance moduli in the individual bands ($\mu^j$ parameter in the eq.~\ref{eq:chi2Phot}) using the extinction law by Cardelli \citep[][]{Cardelli1989_ApJ_345_245} and the E(B-V) values published in T24. Specifically, we adopted the following equations:
\begin{eqnarray}
\rm A_{G_{BP}}=0.504\cdot E(B-V) \\
\rm A_G=0.582\cdot E(B-V) \\
\rm A_{G_{RP}}=0.762\cdot E(B-V)
\end{eqnarray}
to correct the distance moduli $\rm \mu_{G_{BP}}$, $\rm \mu_G$ and $\rm \mu_{G_{RP}}$, obtained from the fitting procedure, and finally, we computed the weighted average of the three values, which we consider as our best estimate of $\mu_0$.

\subsection{Parallax comparison}\label{sec:parallax_comparison}
The reddening free distance moduli, obtained as stated in the previous section, are converted into parallax values (see col. 14 of table D.1 of \href{https://doi.org/10.5281/zenodo.15722385}{Appendix D}).

\begin{figure}
    \centering
    \includegraphics[height=8cm]{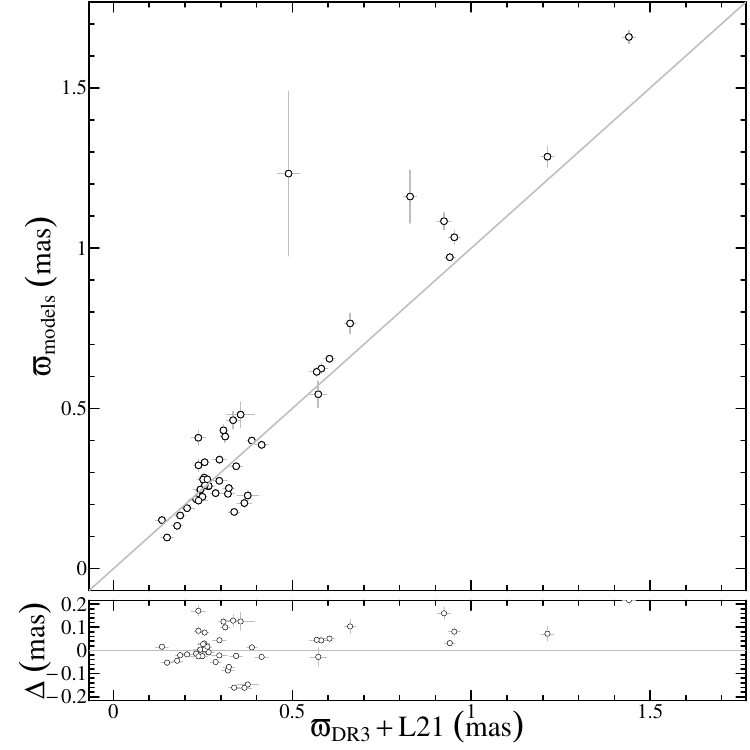}
    \caption{In the top panel, the comparison between the Gaia DR3 parallaxes, corrected according to the L21 recipe, and those obtained from our pulsation model fitting technique is shown together with 1:1 line, represented with the grey solid line.
    In the bottom panel, the residuals, $\rm \Delta = \varpi_{models} - \varpi_{DR3}$, around the 1:1 line, are plotted against the DR3+L21 parallaxes.}
\label{fig:parallaxComparison.L21Corr_fullMwDcepF_Final_singleRun_newSelection_Ebv.lit}
\end{figure}

Figure~\ref{fig:parallaxComparison.L21Corr_fullMwDcepF_Final_singleRun_newSelection_Ebv.lit} shows a comparison between the Gaia DR3 parallax values (including the correction by \citet{Lindegren2021_AA_649_1}, hereinafter L21) and those obtained in this work. 

The analysis presented in the figure indicates that, for parallax values below approximately 0.50 mas, the individual data points show good agreement with the 1:1 line. For parallaxes above this threshold, however, our fitting procedure tends to yield slightly larger values compared to those from DR3+L21. A similar underestimation of parallaxes was reported by \citet{Breuval2020} based on Gaia DR2 data. While the two results are not directly comparable due to differences between DR2 and DR3, the observed behaviour may suggest the persistence of residual systematics at higher parallax values. According to our results, we find an increasing agreement between the two parallax samples as the parallax value decreases. 

To quantify the offset in the Gaia DR3 + L21 parallaxes, we performed a statistical analysis using the theoretical parallaxes as a reference. This analysis involved a series of Monte-Carlo simulations similar to those described in the sec.~\ref{sec:p-factor}.

First, the differences ($\Delta$) between the theoretical parallaxes and those from DR3 were computed. The uncertainty on each $\Delta$ value was determined as the quadrature sum of the uncertainties on the theoretical parallaxes and those on the DR3+L21 parallaxes. Monte Carlo simulations were then carried out by perturbing the $\Delta$ values within their uncertainties, assumed to follow a Gaussian distribution.

The global parallax offset derived from this analysis is 0.014$^{+0.022}_{-0.017}$ mas, with the uncertainty determined using the 5-95\% quantile range. These findings suggest that no significant offset exists in the DR3+L21 parallaxes. A comparable result is obtained by calculating the weighted mean and its associated error from the simulations instead of relying on quantile analysis. In this case, the global parallax offset is 0.005$\pm$0.012 mas, further supporting the conclusion that the L21 correction improves the agreement between the Gaia DR3 parallaxes and the reference values.

Other authors have derived the global offset in the DR3+L21 parallaxes. 
Using eclipsing binaries as reference for distance,  \citet{Stassun2021_ApJL_907_L33} obtained a positive parallax offset equal to +0.015$\pm$0.018 mas, even if even if consistent with zero, in agreement with our findings.

\citet[][]{Wang2022} used a sample of giant stars with accurate known distance and found that the official L21 model largely improves the parallax determination, with a residual offset amounting to +0.0026 mas, (+0.0029 mas) for the five-parameter (six-parameter) solutions.

\citet{Fabricius2021} compared the EDR3 parallaxes with the distances obtained from other external catalogues (see their Tab.1) and found that the L21 correction significantly improves the parallax comparison, in agreement with our results.

Significant negative global offset values were found by other authors in the sense that L21 recipe seems to over-correct the parallaxes, returning parallaxes larger than the reference values. For instance, \citet{Riess2021_ApJL_908_L6} found a global offset  $\Delta \varpi=-0.014 \pm 0.006$ mas, while two other recent studies, \citet{Molinaro2023_mnras_520_4154, Bhardwaj2021_ApJ_909_200} found offsets  $\Delta \varpi=-0.022 \pm 0.004$ mas and $\Delta \varpi=-0.022 \pm 0.002$ mas, respectively. 

We refer the reader to Fig.10 in \citet{Molinaro2023_mnras_520_4154} for a more complete review of the most recent determinations of the global parallax offset.

\subsection{Period-Wesenheit-Metallicity relation}
To minimize the effect of reddening on the distance determination, the reddening--free by definition Period-Wesenheit (PW) relations and their metal-dependent versions (PWZ) are often adopted (see e.g. T24).
The period and metallicity values characterizing the investigated Cepheid sample, are not representative of the entire expected range for CCs. As a consequence, we did not fit the PWZ relation by using the mean parameters of the selected variables, but we studied how they distribute in comparison with the PWZ relation recently obtained by our group within the C-MetaLL project \citep[see e.g][]{Ripepi2021_mnras_508_4047, Trentin2024_AA_681_A65}.

\begin{figure}
    \centering
    \includegraphics[trim={0cm 5.9cm 0.0cm 0cm},clip, height=5cm]{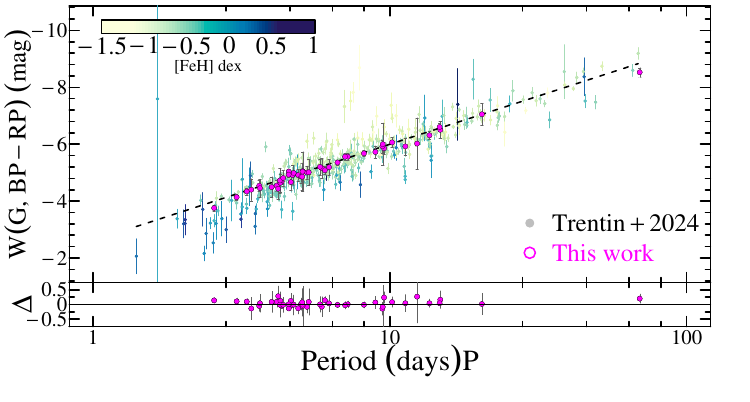}
    \caption{Absolute Wesenheit magnitude is plotted against against the pulsation period. The Fundamental CCs from T24 are plotted, by using a coloured scale according to their [FeH] values, together with their best PWZ fitted relation (dashed line). The same quantities for the sample studied in this work are also plotted, together with their errors, by using magenta circles. Their residuals around the plotted fit, are also shown in the bottom panel. }
\label{fig:PeriodAbl_fullMwDcepF_Final_singleRun_newSelection}
\end{figure}

To this end, we calculated the absolute mean Wesenheit magnitude\footnote{Despite the inversion of parallax for the calculation of distance, and consequently of absolute magnitude, introducing a bias and leading to the loss of the Gaussian nature of the parallax error \citep{Feast1997_mnras_286_L1, Arenou1999_ASPC_167_13}, this test is purely indicative, aimed at verifying how well our sample follows the PWZ relation.} in the Gaia bands using the parallax corrections from L21 and the mean magnitudes $G$, $G_{BP}$, and $G_{RP}$ from the sample of CCs by T24. The obtained values were plotted as a function of period in the top panel of Fig.~\ref{fig:PeriodAbl_fullMwDcepF_Final_singleRun_newSelection},  by adopting a colour scale showing the trend of the [FeH] values. The data representing our sample are also displayed with magenta circles. Additionally, the best-fit PWZ relation from T24 (see their Table 3) is plotted as a dashed black line. The residuals of our sample relative to the best-fit relation are shown in the bottom panel of the same figure. The agreement appears very good. A statistical analysis hints that there is a  marginal systematic residual with median value equal to 0.016 mag, compared with the fitted line. Moreover, the robust standard deviation of the distribution of the residuals amounts to 0.10 mag, indicating that our sample is reasonably distributed around the PWZ relation by T24.

\subsection{Mass-Luminosity relation}

\begin{figure}
    \centering
    \includegraphics[height=9cm]{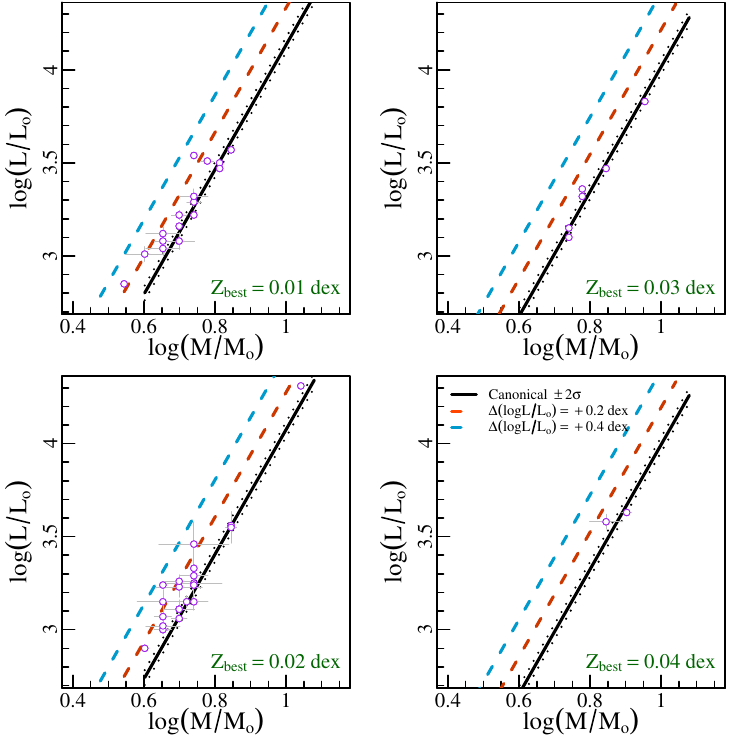}
    \caption{The mass and luminosity best values obtained by our pulsation model fit technique, are plotted for all the selected sources (violet points), together with the expected canonical ML relation (black solid line), the mild (+0.2 dex) non-canonical ML relation (red  dashed line) and the full (+0.4 dex) non-canonical ML relation (blue dashed  line).}
    \label{fig:ML.relation_fullMwDcepF_Final_singleRun_newSelection}
\end{figure}

The inferred individual mass and luminosity values allow us to understand how the best fit models compare with the ML relations predicted by stellar evolution.
In Fig.~\ref{fig:ML.relation_fullMwDcepF_Final_singleRun_newSelection} we show the distribution of the obtained best-fit models (violet circles), divided in the four selected metallicity bins, in the ML plane, compared with the relations predicted by the adopted stellar evolution scenario in the canonical assumption (black solid line),  including a mild (+0.2 dex, red dashed line) or full (+0.4 dex, blue dashed line) over-luminosity due to overshooting or mass loss, or rotation.
We notice that for each metallicity bin, the bulk of the model results are located within the canonical and the mildly non canonical relations, with a smaller number of models between the mildly and the fully non-canonical lines.
The inferred dispersion supports the idea that the observed Cepheid sample do not obey a unique universal ML relation, possibly reflecting the stochastic nature of mass loss, thus displaying a non-negligible spread in the ML plane. This dispersion is also expected to affect the coefficients of the PL, PLC and PW relations.

\section{Conclusions}\label{sec:conclusions}
In this work, we used hydrodynamic pulsation models to fit a sample of Galactic CCs for which photometry, radial velocity, and parallax data from the Gaia mission's DR3 were available, complemented with recent high-resolution spectroscopic metallicities. Several criteria were considered for the sample selection: i) the widest possible coverage of the expected period range for CCs; ii) well-sampled light and radial velocity curves in phase; iii) sufficiently accurate parallaxes ($\sigma_\varpi/\varpi < 50\%$); iv) metallicities obtained through recent high-resolution spectroscopic measurements.

Using the models generated by our hydrodynamic pulsation code, we performed a fitting procedure based on the minimization of the $\chi^2$ function. By employing a set of bootstrap Monte Carlo simulations, we investigated how the best-fit model varied by randomly resampling the input data. This procedure allowed us to determine the uncertainties on both the fit parameters and the structural parameters estimated for each variable.

The main results of this best-fitting method are:
\begin{itemize}
\item The pulsational periods are accurately reproduced by hydrodynamical models, with percentage differences that are not larger than 10\%.
\item No significant trend was found for the inferred p-factor as a function of the pulsation period,although the relatively large uncertainties associated with several p-factor estimates may hinder the detection of a possible underlying trend. By averaging the obtained values, our best estimate of the p-factor is  $p = 1.22 \pm 0.05$.  Moreover , we must extend the investigated sample by uniformly populating the period ranges to provide more stringent results.

\item  The parallaxes obtained from the inferred de-reddened distance moduli are in satisfactory agreement with the Gaia DR3 parallaxes corrected according to the L21 recipe, with a residual global offset that is consistent with zero.

\item The inferred parallaxes also provide a very good agreement with the PWZ relation by T24.

\item The stellar masses and luminosities of the derived best-fit model solutions suggest that, for each selected metallicity bin, the bulk of the investigated Cepheids obey an ML relation that is between the canonical and the mildly over-luminous (+0.2 dex) one, with a smaller number of pulsators displaying a brighter relation.
\end{itemize}

The results of the presented analysis confirm the global predictive capability of the adopted theoretical scenario. We plan to better investigate residual uncertainties, e.g. the ones connected to the reddening or p--factor determination, by extending both the observational sample and parameter space of the adopted model set, also benefiting from upcoming facilities devoted to variability studies, such as the Rubin LSST.

\section{Data availability}
\href{https://doi.org/10.5281/zenodo.15722385}{Appendix D}

\begin{acknowledgements}
The Authors thank the Referee, for his detailed analysis of  this work, that  impressively improved its contents and readability.
\end{acknowledgements}

% WARNING
%-------------------------------------------------------------------
% Please note that we have included the references to the file aa.dem in
% order to compile it, but we ask you to:
%
% - use BibTeX with the regular commands:
\bibliographystyle{aa} % style aa.bst
%\bibliography{references_proModelFitting} % your references Yourfile.bib

%
% - join the .bib files when you upload your source files
%-------------------------------------------------------------------

\begin{appendix}
    \section{Initial Fourier model}\label{app:initial-fourier-model}
After folding the input data with the observed pulsational period, the fitting routine constructs a Fourier model for each input time series of photometric and radial velocity data. 

The routine uses some useful information provided by this initial Fourier model. In particular, the phase difference between the maximum of light (minimum of velocity) of the observed time series and that of the theoretical curves, gives a first rough guess of the phase shift to be applied to the model in order to fit observations. Furthermore, the Fourier peak-to-peak amplitude is used by the fitting routine to compute the $\chi^2_{Amp}$ term of eq.~\ref{eq:chi2Phot} and \ref{eq:chi2RV}.

\section{The PTLM  and ML equations}\label{app:ML-PTLM equations}
The PTLM relation adopted in this work is from \citet{DeSomma2020JSS_247_30}:
\begin{equation}\label{eq:ptlm}
\rm  \log P = a_1 + a_2\log T_{eff} + a_3\log(M/M_\odot) + a_4(\log(L/L_\odot) + \Delta_{ML})
\end{equation}
where the coefficients $a_i$ depend on the selected $\alpha$ value (see Tab.~\ref{tab:ptlm-ml-coeffs}), while the $\Delta_{ML}$ takes into account the over-luminosity level with respect to the canonical ML relation ($\Delta_{ML}=0$), having the following form \citep{Bono_2000_ApJ_543_955}:
\begin{equation}\label{eq:ml}
    \rm \log L = b_1 + b_2 \log M + b_3  \log Z + b_4\log Y.
\end{equation}
where the coefficients are given in Tab.~\ref{tab:ptlm-ml-coeffs}.

A generic model is defined by its elemental composition (Z, Y), the pulsational period (P), the effective temperature ($\rm T_{eff}$), the mass (M), and the luminosity ($\rm \log L/L_\odot$).
Fixing the elemental composition, for every mass (M) and effective temperature values ($\rm T_{eff}$), the luminosity of the model is given by summing the canonical value, obtained from eq.~\ref{eq:ml}, to the non-canonical correction $\Delta_{ML}$ given by:
\begin{equation}
\begin{aligned}
       \Delta_{ML} = \frac{1}{a_4}\Big (\log P - a_1 - a_2 \log T_{eff} -(a_3 + a_4b_2)\log (M/M_\odot) \\ - a_4b_3\log Y -a_4b_4 \log Z - a_4b_1 \Big )
       \end{aligned}
       \label{eq:deltaML}
\end{equation}
such that the computed theoretical period matches the value obtained from the observations. 

We note here that, we did not consider the PTLM coefficient errors in our computation because they affect the inferred parameters by amounts that are significantly smaller than the adopted steps in the corresponding model grid.

\begin{table*}
    \centering
    \caption{The coefficients of the PTLM form \citet{DeSomma2020JSS_247_30} and of the ML relation \citet{Bono_2000_ApJ_543_955} we used in the current work.}
        \centering
        \begin{tabular}{ccccc}
            \hline\hline             
              \multicolumn{5}{c}{PTML relation}\\
    \hline
    $\alpha$ & $\rm a_1$ & $\rm a_2$ & $\rm a_3$ & $\rm a_4$\\
    \hline
  1.5 & 10.268$\pm$0.001 & -3.192$\pm$0.025 & -0.758$\pm$0.015 & 0.919$\pm$0.005\\
 1.7 & 10.538$\pm$0.002 & -3.258$\pm$0.002 & -0.749$\pm$0.019 & 0.911$\pm$0.007\\
 1.9 & 11.488$\pm$0.003 & -3.469$\pm$0.089 & -0.695$\pm$0.012 & 0.847$\pm$0.006\\
            \hline
        \end{tabular}
        \centering
        \begin{tabular}{ccccc}
  \multicolumn{5}{c}{ML relation}\\
         \hline
  &       $\rm b_1$ & $\rm b_2$ & $\rm b_3$ & $\rm b_4$\\
  \hline
 &0.90$\pm$0.02 & +3.35$\pm$0.03 & 1.36$\pm$0.13 & -0.34$\pm$0.02\\
            \hline
        \end{tabular}
    \label{tab:ptlm-ml-coeffs} 
\end{table*}

\section{Minimization of the $\chi^2$}\label{app-chi2-minimization}
Requiring that the first derivatives of the eq.~\ref{eq:chi2Phot} and eq.~\ref{eq:chi2RV} with respect to the parameters of the fit, are equal to zero, it is possible to find the following equations:
%\newpage 

\begin{strip}
  \begin{equation}
    \begin{split}
  -\sum_{i=1}^{N^j_{pts}}\left(\frac{m_i^j(\phi^j_i)\dot{M}^j\left(\phi^j_i+\delta \Phi^j\right)}{\left(\sigma^j_i\right)^2} \right) +
  \sum_{i=1}^{N^j_{pts}}\left(\frac{M^j\left(\phi^j_i+\delta \Phi^j\right) \dot{M}^j\left(\phi^j_i+\delta \Phi^j\right)  }{\left(\sigma^j_i\right)^2} \right) +
  \frac{1}{B}\left(\sum_{i=1}^{N^j_{pts}}\left(\frac{M^j\left(\phi^j_i+\delta \Phi^j\right)}{\left(\sigma^j_i\right)^2} \right) - A \right)\sum_{i=1}^{N^j_{pts}}\left(\frac{\dot{M}^j\left(\phi^j_i+\delta \Phi^j\right)}{\left(\sigma^j_i\right)^2} \right) = 0
  \end{split}\label{eq:chi-deltaphi-phot}
    \end{equation}
%\newpage

 \begin{equation}
    \begin{split}\label{eq:chi-deltaphi-rv}    
   C\sum_{l=1}^{M_{pts}}\left(\frac{V_{puls}\left(\phi_l + \delta \Phi_{RV}\right)}{\sigma_l^{RV}}\right)^2 \sum_{l=1}^{M_{pts}}\left(\frac{v_l\dot{V}_{puls}\left(\phi_l + \delta \Phi_{RV}\right)}{\left(\sigma_l^{RV}\right)^2}\right) - \left(\sum_{l=1}^{M_{pts}}\left(\frac{V_{puls}\left(\phi_l + \delta \Phi_{RV}\right)}{\left(\sigma^{RV}_l\right)^2}\right)\right)^2\sum_{l=1}^{M_{pts}}\left(\frac{v_l\dot{V}_{puls}\left(\phi_l + \delta \Phi_{RV}\right)}{\left(\sigma_l^{RV}\right)^2}\right) \\
 -C\sum_{l=1}^{M_{pts}}\left(\frac{v_lV_{puls}\left(\phi_l + \delta \Phi_{RV}\right)}{\left(\sigma_l^{RV}\right)^2}\right)\sum_{l=1}^{M_{pts}}\left(\frac{\dot{V}_{puls}\left(\phi_l + \delta \Phi_{RV}\right)V_{puls}\left(\phi_l + \delta \Phi_{RV}\right)}{\left(\sigma_l^{RV}\right)^2}\right) +\\
 + A'\sum_{l=1}^{M_{pts}}\left(\frac{V_{puls}\left(\phi_l + \delta \Phi_{RV}\right)}{\left(\sigma^{RV}_l\right)^2}\right)\sum_{l=1}^{M_{pts}}\left(\frac{\dot{V}_{puls}\left(\phi_l + \delta \Phi_{RV}\right)V_{puls}\left(\phi_l + \delta \Phi_{RV}\right)}{\left(\sigma_l^{RV}\right)^2}\right) +\\
  -A'\sum_{l=1}^{M_{pts}}\left(\frac{V_{puls}\left(\phi_l + \delta \Phi_{RV}\right)}{\sigma_l^{RV}}\right)^2\sum_{l=1}^{M_{pts}}\left(\frac{\dot{V}_{puls}\left(\phi_l + \delta \Phi_{RV} \right)}{\left(\sigma_l\right)^2} \right) + \\
 +\sum_{l=1}^{M_{pts}}\left(\frac{V_{puls}\left(\phi_l + \delta \Phi_{RV}\right)}{\left(\sigma^{RV}_l\right)^2}\right)\sum_{l=1}^{M_{pts}}\left(\frac{v_lV_{puls}\left(\phi_l + \delta \Phi_{RV}\right)}{\left(\sigma_l^{RV}\right)^2}\right)\sum_{l=1}^{M_{pts}}\left(\frac{\dot{V}_{puls}\left(\phi_l + \delta \Phi_{RV} \right)}{\left(\sigma_l\right)^2} \right) = 0 
    \end{split}
    \end{equation}
\end{strip}

where $A=\sum_{i=1}^{N^j_{pts}} \frac{m_i^j}{\left(\sigma_i^j \right)^2}$, $B=\sum_{i=1}^{N^j_{pts}}\frac{1}{\left(\sigma_i^j \right)^2}$, $C=\sum_{l=1}^{M_{pts}}\frac{1}{\left(\sigma_l^{RV} \right)^2}$,
$A'=\sum_{l=1}^{M_{pts}} \frac{v_l}{\left(\sigma^{RV}_l \right)^2}$, while the unknown parameters are the two phase shifts $\delta \Phi^j$ and $\delta \Phi_{RV}$.

After solving the equations for the two unknown phase shifts, the apparent distance modulus in the j-th band is given by: 
\begin{equation}\label{eq:chi2-mui}
\rm \mu^j=\frac{1}{B}\cdot \left (\frac{\sum_{i=1}^{N^j_{pts}}M^j(\Phi_i^j+\delta \Phi^j)}{\sigma_i^2} - A\right )    
\end{equation},
while the barycentric $\gamma$ velocity and the projection factor can be obtained from the following equations:
\begin{equation}\label{eq:chi2-gamma}
    \rm \gamma = \frac{A'\cdot(E-F) + H\cdot B' - D\cdot B'}{B'\cdot G - B'^2 -C\cdot F + C\cdot E},
\end{equation}
\begin{equation}\label{eq:chi2-pfact}
    \rm 1/p = \frac{A'- \gamma \cdot C}{B},
\end{equation}
where $B'= \sum_{l=1}^{M_{pts}}\frac{V_{puls}(\phi_l+\delta \Phi_{RV})}{(\sigma_l^{RV})^2}$, $D=\sum_{l=1}^{M_{pts}}\left (\frac{v_l\cdot V_{puls}(\phi_l + \delta \Phi_{RV})}{(\sigma_l^{RV})} \right )$, $E=\sum_{l=1}^{M_{pts}}\frac{V_{puls}(\phi_l+\delta \Phi_{RV})}{\sigma_l^{RV}}$, $F=\sum_{l=1}^{M_{pts}}\frac{V_{puls}(\phi_l+\delta \Phi_{RV})\cdot \dot{V}(\phi_l+\delta \Phi_{RV})}{\sigma_l^2}$, $G=\sum_{l=1}^{M_{pts}}\frac{\dot{V}_{puls}(\phi_l + \delta \Phi_{RV})}{\sigma_l^2}$ and $H=\sum_{l=1}^{M_{pts}}\frac{v_l\cdot \dot{V}_{puls}(\phi_l + \delta \Phi_{RV})}{\sigma_l^2}$.
Despite, their long and complicated expression, equations ~\ref{eq:chi2-mui}, \ref{eq:chi2-gamma} and \ref{eq:chi2-pfact} consist of summations and are very fast to be computed, once the $\delta \Phi$s are obtained from the solution of the equations \ref{eq:chi-deltaphi-rv}.

\section{Best models light and radial velocity curves}\label{app:bestModelsPlot}
Figure~\ref{fig:AllBestModels_fullMwDcepF_Final_singleRun_newSelection} shows the best models, obtained by fitting both photometric and radial velocity curves. The best-fit parameters for the 46 stars of our final sample are contained in the Tab.~\ref{tab:bestFitParams}.

\begin{figure*}
  \centering
\includegraphics[page=1, height=19cm]{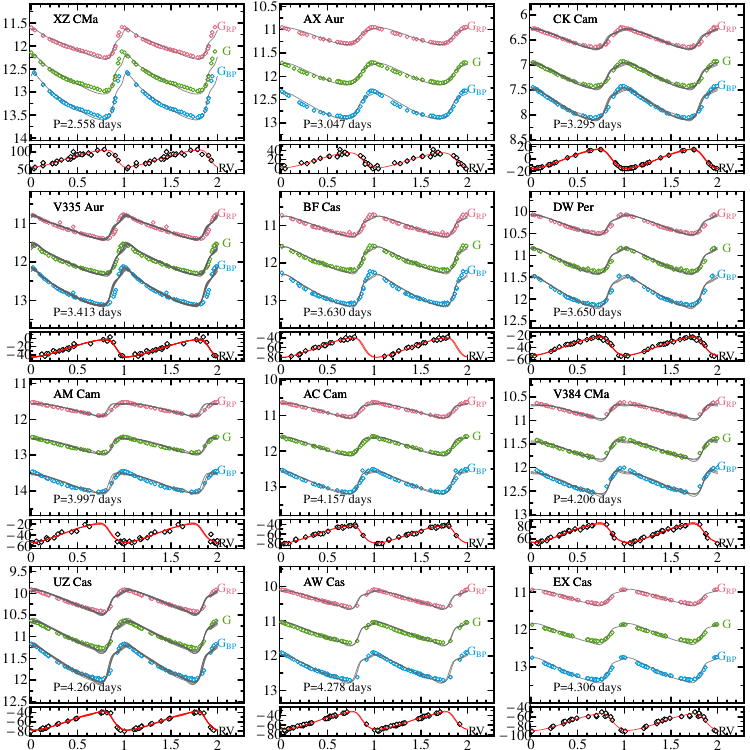}
   \caption{The model fitting results obtained by using both the photometric and radial velocity data. In the top panels, the $G_{BP}$, $G$ and $G_{RP}$ data are plotted by using different colours, while the best models, obtained from our Monte-Carlo bases procedure, are shown as solid gray lines. In the bottom panels we plot the best fit results for the radial velocity data: the observations are plotted as empty black rhombuses, while solid red lines represent the best models obtained from Monte-Carlo simulations}\label{fig:AllBestModels_fullMwDcepF_Final_singleRun_newSelection}
\end{figure*}
\addtocounter{figure}{-1}
\begin{figure*}[htbp]
  \centering
  \includegraphics[page=2, height=19cm]{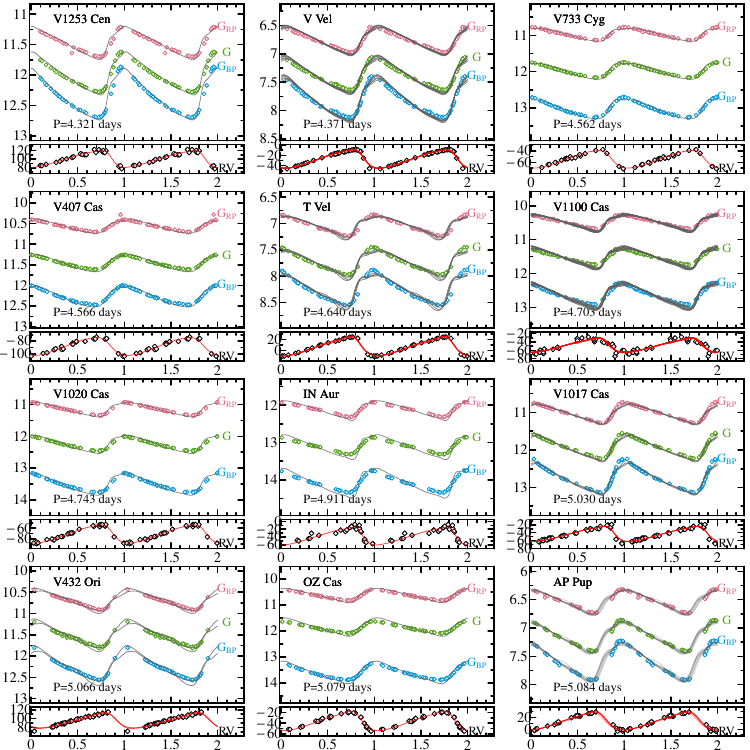}
\caption{Continued}
\end{figure*}
\addtocounter{figure}{-1}
\begin{figure*}[htbp]
  \centering
  \includegraphics[page=3, height=19cm]{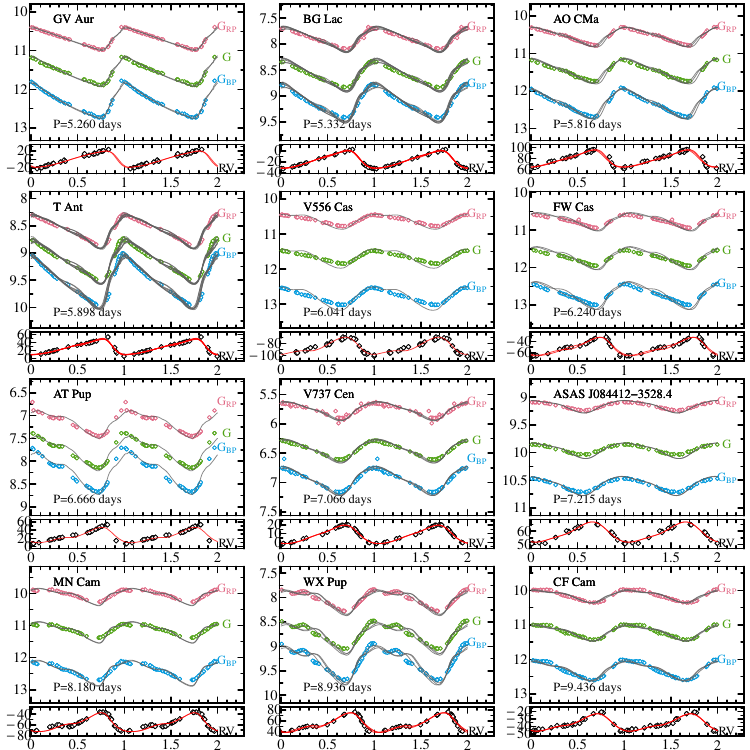}
\caption{Continued}
\end{figure*}
\addtocounter{figure}{-1}
\begin{figure*}[htbp]
  \centering
  \includegraphics[page=4, height=19cm]{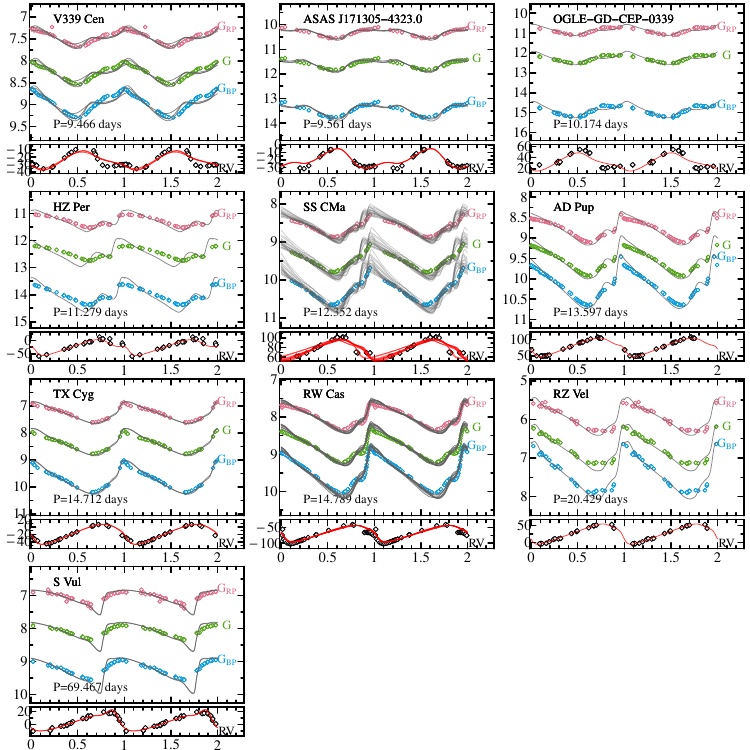}
\caption{Continued}
\end{figure*}

\begin{table*}
\centering
\caption{Summary of the best-fit parameters for the 46 stars in the final studied sample. } 
\footnotesize\setlength{\tabcolsep}{3.0pt}
\begin{tabular}{llllllllllllllc}
  \hline
star & $\rm P_{obs}$ & $\rm P_{best}$ & $\alpha$ & $\rm [FeH]$ &  $\rm Z_{best}$ & $\rm Y_{best}$ & M & $\rm T_{eff}$ & $\rm \log(L/L_\odot)$ & $\Delta_{ML}$ & p & $\gamma$ & $\varpi$ & $\chi^2$\\ 
 & (days) & (days) &  & (dex) &  (dex) & (dex) & $(\rm M_\odot)$ & (K) & (dex) & (dex) &  & (km/s) & (mas) & \\ 
  \hline
XZ CMa & 2.5578 & 2.4946 & 1.70 & -0.26 &  0.01 & 0.26 & $3.5 _{-0.5}^{+0.5}$ & $6050 _{-50}^{+50}$ & $2.85 _{-0.05}^{+0.05}$ & $0.25 _{-0.15}^{+0.15}$ & $1.18_{-0.19}^{+0.22}$ & $82 _{-2}^{+2}$ & $0.134 _{-0.002}^{+0.002}$ &$2.98^{-0.27}_{+0.37}$\\ 
  AX Aur & 3.0468 & 3.0446 & 1.70 & -0.08 & 0.02 & 0.28 & $4.0 _{-0.5}^{+0.5}$ & $5750 _{-50}^{+50}$ & $2.90 _{-0.04}^{+0.04}$ & $0.16 _{-0.15}^{+0.15}$ & $1.17_{-0.20}^{+0.20}$ & $19 _{-1}^{+1}$ & $0.247 _{-0.005}^{+0.005}$ & $1.41^{-0.39}_{+0.37}$\\ 
  CK Cam & 3.2949 & 3.3004 & 1.65 & 0.06 & 0.02 & 0.28 & $4.5 _{-0.75}^{+0.75}$ & $5750 _{-25}^{+25}$ & $2.98 _{-0.07}^{+0.07}$ & $0.07 _{-0.19}^{+0.19}$ & $1.33_{-0.07}^{+0.07}$ & $-0.2 _{-0.6}^{+0.6}$ & $1.659 _{-0.020}^{+0.020}$ & $1.53^{-1.01}_{+0.84}$ \\ 
  V335 Aur & 3.4127 & 3.3152 & 1.70 & -0.33 & 0.01 & 0.26 & $5.0 _{-0.5}^{+0.5}$ & $5950 _{-50}^{+50}$ & $3.08 _{-0.05}^{+0.05}$ & $-0.04 _{-0.13}^{+0.13}$ & $1.25_{-0.30}^{+0.30}$ & $-22 _{-3}^{+3}$ & $0.217 _{-0.002}^{+0.002}$ & $1.12^{-0.38}_{+0.41}$\\ 
  BF Cas & 3.6302 & 3.6619 & 1.50 & -0.08 & 0.02 & 0.28 & $5.0 _{-0.25}^{+0.25}$ & $5700 _{-50}^{+50}$ & $3.06 _{-0.02}^{+0.02}$ & $-0.01 _{-0.06}^{+0.06}$ & $1.30_{-0.12}^{+0.14}$ & $-59 _{-4}^{+3}$ & $0.258 _{-0.007}^{+0.007}$ & $0.97^{-0.25}_{0.28}$\\ 
  DW Per & 3.6502 & 3.6472 & 1.60 & -0.08 &  0.02 & 0.28 & $4.5 _{-0.5}^{+0.5}$ & $5700 _{-50}^{+50}$ & $3.02 _{-0.04}^{+0.04}$ & $0.11 _{-0.12}^{+0.12}$ & $1.35_{-0.16}^{+0.18}$ & $-37 _{-1}^{+1}$ & $0.340 _{-0.009}^{+0.009}$ & $1.35^{-0.40}_{0.52}$\\ 
  AM Cam & 3.9972 & 3.9081 & 1.60 & -0.16 & 0.01 & 0.26 & $4.5 _{-0.5}^{+0.5}$ & $5600 _{-50}^{+50}$ & $3.04 _{-0.04}^{+0.04}$ & $0.06 _{-0.12}^{+0.12}$ & $1.26_{-0.35}^{+0.31}$ & $-36 _{-4}^{+5}$ & $0.246 _{-0.009}^{+0.009}$ & $1.19^{-0.31}_{0.28}$\\ 
  AC Cam & 4.1566 & 4.0160 & 1.70 & -0.16 &  0.01 & 0.26 & $4.5 _{-0.5}^{+0.5}$ & $5700 _{-50}^{+50}$ & $3.08 _{-0.02}^{+0.02}$ & $0.11 _{-0.06}^{+0.06}$ & $1.32_{-0.12}^{+0.14}$ & $-60 _{-1}^{+1}$ & $0.320 _{-0.008}^{+0.008}$ & $0.96^{-0.33}_{0.38}$\\ 
  V384 CMa & 4.2060 & 4.0786 & 1.60 & -0.04 & 0.01 & 0.26 & $4.0 _{-0.5}^{+0.5}$ & $5550 _{-50}^{+50}$ & $3.01 _{-0.04}^{+0.04}$ & $0.20 _{-0.12}^{+0.12}$ & $1.24_{-0.21}^{+0.21}$ & $69 _{-2}^{+2}$ & $0.257 _{-0.005}^{+0.005}$ & $0.95^{-0.36}_{0.38}$\\ 
  UZ Cas & 4.2596 & 4.2491 & 1.50 & -0.08 & 0.02 & 0.28 & $5.0 _{-0.5}^{+0.5}$ & $5600 _{-50}^{+50}$ & $3.11 _{-0.05}^{+0.05}$ & $0.04 _{-0.10}^{+0.10}$ & $1.19_{-0.15}^{+0.14}$ & $-59 _{-2}^{+2}$ & $0.278 _{-0.005}^{+0.005}$ & $16.95^{-1.85}_{1.76}$\\ 
  AW Cas & 4.2782 & 4.2689 & 1.50 & 0.00 & 0.02 & 0.28 & $4.5 _{-0.5}^{+0.5}$ & $5600 _{-50}^{+50}$ & $3.07 _{-0.02}^{+0.02}$ & $0.16 _{-0.06}^{+0.06}$ & $1.22_{-0.12}^{+0.12}$ & $-51 _{-4}^{+2}$ & $0.386 _{-0.012}^{+0.012}$ & $2.58^{-0.39}_{0.42}$\\ 
  EX Cas & 4.3055 & 4.2914 & 1.60 & -0.10 &  0.02 & 0.28 & $5.0 _{-0.5}^{+0.5}$ & $5600 _{-50}^{+50}$ & $3.11 _{-0.04}^{+0.04}$ & $0.04 _{-0.12}^{+0.12}$ & $1.17_{-0.21}^{+0.23}$ & $-71 _{-2}^{+2}$ & $0.224 _{-0.004}^{+0.004}$ & $8.83^{-0.57}_{0.71}$\\ 
  V1253 Cen & 4.3209 & 4.1646 & 1.70 & -0.26 & 0.01 & 0.26 & $5.0 _{-0.5}^{+0.5}$ & $5800 _{-50}^{+50}$ & $3.16 _{-0.03}^{+0.03}$ & $0.04 _{-0.12}^{+0.12}$ & $1.31 _{-0.15}^{+0.14}$ & $99 _{-1}^{+1}$ & $0.0973 _{-0.0002}^{+0.0002}$ & $1.14^{-0.28}_{0.29}$\\ 
  V Vel & 4.3711 & 4.3647 & 1.55 & 0.00 &  0.02 & 0.28 & $5.3 _{-0.5}^{+0.5}$ & $5650 _{-50}^{+50}$ & $3.15 _{-0.03}^{+0.03}$ & $0.01 _{-0.07}^{+0.07}$ & $1.25_{-0.10}^{+0.10}$ & $-26 _{-1}^{+1}$ & $1.034 _{-0.021}^{+0.021}$  & $6.30^{-10.11}_{10.41}$\\ 
  V733 Cyg & 4.5621 & 4.6508 & 1.50 & 0.14 & 0.03 & 0.28 & $5.5 _{-0.5}^{+0.5}$ & $5350 _{-50}^{+50}$ & $3.10 _{-0.03}^{+0.03}$ & $-0.04 _{-0.10}^{+0.10}$ & $1.21_{-0.13}^{+0.14}$ & $-54 _{-2}^{+2}$ & $0.323 _{-0.019}^{+0.019}$ & $1.14^{-0.38}_{0.37}$ \\ 
  V407 Cas & 4.5661 & 4.5533 & 1.65 & 0.08 &  0.02 & 0.28 & $5.5 _{-0.5}^{+0.5}$ & $5550 _{-50}^{+50}$ & $3.16 _{-0.04}^{+0.04}$ & $-0.049 _{-0.011}^{+0.011}$ & $1.21_{-0.22}^{+0.22}$ & $-88 _{-1}^{+1}$ & $0.279 _{-0.009}^{+0.009}$ & $3.39^{-0.54}_{0.51}$\\ 
  T Vel & 4.6400 & 4.4575 & 1.70 & -0.16 & 0.01 & 0.26 & $4.5 _{-0.5}^{+0.5}$ & $5650 _{-50}^{+50}$ & $3.12 _{-0.04}^{+0.04}$ & $0.15 _{-0.13}^{+0.13}$ & $1.26_{-0.08}^{+0.07}$ & $5 _{-1}^{+1}$ & $0.972 _{-0.003}^{+0.003}$ & $6.28^{-8.25}_{8.66}$\\ 
  V1100 Cas & 4.7026 & 4.7076 & 1.50 & -0.01 & 0.02 & 0.28 & $5.5 _{-0.5}^{+0.5}$ & $5450 _{-100}^{+100}$ & $3.15 _{-0.05}^{+0.05}$ & $-0.05 _{-0.14}^{+0.14}$ & $1.25 _{-0.32}^{+0.34}$ & $-49 _{-3}^{+3}$ & $0.233 _{-0.012}^{+0.012}$ & $8.66^{-4.30}_{4.29}$\\ 
  V1020 Cas & 4.7429 & 4.8059 & 1.50 & 0.12 & 0.03 & 0.28 & $5.5 _{-0.5}^{+0.5}$ & $5450 _{-50}^{+50}$ & $3.15 _{-0.02}^{+0.02}$ & $0.01_{-0.11}^{+0.11}$ & $1.16_{-0.08}^{+0.09}$ & $-71 _{-1}^{+1}$ & $0.177 _{-0.011}^{+0.011}$ & $6.01^{-0.24}_{0.31}$\\ 
  IN Aur & 4.9107 & 4.7611 & 1.60 & -0.31 & 0.01 & 0.26 & $5.5 _{-0.5}^{+0.5}$ & $5650 _{-50}^{+50}$ & $3.22 _{-0.02}^{+0.02}$ & $-0.05 _{-0.10}^{+0.10}$ & $1.17_{-0.29}^{+0.26}$ & $-38 _{-4}^{+3}$ & $0.152 _{-0.005}^{+0.005}$ & $14.26^{-0.36}_{0.38}$\\ 
  V1017 Cas & 5.0297 & 4.7908 & 1.70 & -0.21 & 0.01 & 0.26 & $5.0 _{-0.5}^{+0.5}$ & $5750 _{-50}^{+50}$ & $3.22 _{-0.03}^{+0.03}$ & $0.10 _{-0.05}^{+0.05}$ & $1.25_{-0.14}^{+0.15}$ & $-43 _{-1}^{+1}$ & $0.166 _{-0.004}^{+0.004}$ & $2.06^{-0.84}_{0.83}$\\ 
  V432 Ori & 5.0658 & 4.9938 & 1.50 & -0.13 & 0.02 & 0.28 & $4.5 _{-0.75}^{+0.75}$ & $5600 _{-50}^{+50}$ & $3.15 _{-0.07}^{+0.07}$ & $0.24 _{-0.14}^{+0.14}$ & $1.14_{-0.16}^{+0.17}$ & $94 _{-4}^{+6}$ & $0.213 _{-0.002}^{+0.002}$ & $3.30^{-0.80}_{0.84}$\\ 
  OZ Cas & 5.0792 & 5.0360 & 1.50 & 0.03 &  0.02 & 0.28 & $5.5 _{-0.5}^{+0.5}$ & $5650 _{-50}^{+50}$ & $3.24 _{-0.04}^{+0.04}$ & $0.03 _{-0.10}^{+0.10}$ & $1.16 _{-0.12}^{+0.12}$ & $-37 _{-1}^{+1}$ & $0.544 _{-0.041}^{+0.041}$ & $22.78^{-0.54}_{0.45}$\\ 
  AP Pup & 5.0838 & 5.0330 & 1.60 & -0.02 & 0.02 & 0.28 & $5.5 _{-0.5}^{+0.5}$ & $5600 _{-75}^{+75}$ & $3.22 _{-0.03}^{+0.03}$ & $0.02 _{-0.12}^{+0.12}$ & $1.23 _{-0.09}^{+0.09}$ & $12 _{-1}^{+1}$ & $1.084 _{-0.026}^{+0.026}$ & $7.37^{-5.28}_{7.24}$\\ 
  GV Aur & 5.2599 & 5.0287 & 1.70 & -0.24 & 0.01 & 0.26 & $5.5 _{-0.5}^{+0.5}$ & $5800 _{-50}^{+50}$ & $3.29 _{-0.03}^{+0.03}$ & $0.03 _{-0.04}^{+0.04}$ & $1.20_{-0.08}^{+0.09}$ & $0 _{-3}^{+2}$ & $0.189 _{-0.0018}^{+0.0018}$ & $1.13^{-0.33}_{0.35}$\\ 
  BG Lac & 5.3317 & 5.2823 & 1.55 & 0.04 &  0.02 & 0.28 & $5.5 _{-0.5}^{+0.5}$ & $5500 _{-75}^{+75}$ & $3.22 _{-0.06}^{+0.06}$ & $0.03 _{-0.09}^{+0.09}$ & $1.21 _{-0.14}^{+0.15}$ & $-16 _{-1}^{+1}$ & $0.624 _{-0.016}^{+0.016}$ & $5.77^{-3.22}_{3.81}$\\ 
  AO CMa & 5.8157 & 5.7479 & 1.50 & 0.01 & 0.02 & 0.28 & $5.5 _{-0.5}^{+0.5}$ & $5600 _{-50}^{+50}$ & $3.29 _{-0.04}^{+0.04}$ & $0.082 _{-0.10}^{+0.10}$ & $1.19 _{-0.12}^{+0.14}$ & $78 _{-2}^{+2}$ & $0.236 _{-0.005}^{+0.005}$ & $4.98^{-1.00}_{1.28}$\\ 
  T Ant & 5.8984 & 5.6211 & 1.60 & -0.23 &  0.01 & 0.26 & $5.5 _{-0.5}^{+0.5}$ & $5700 _{-50}^{+50}$ & $3.32 _{-0.04}^{+0.04}$ & $0.07 _{-0.11}^{+0.11}$ & $1.16_{-0.12}^{+0.13}$ & $28 _{-4}^{+4}$ & $0.412 _{-0.018}^{+0.018}$ & $24.36^{-6.49}_{5.30}$\\ 
  V556 Cas & 6.0411 & 5.9490 & 1.40 & -0.04 &  0.02 & 0.28 & $4.5 _{-0.5}^{+0.5}$ & $5600 _{-50}^{+50}$ & $3.24 _{-0.05}^{+0.05}$ & $0.32 _{-0.12}^{+0.12}$ & $1.14 _{-0.27}^{+0.27}$ & $-86 _{-1}^{+1}$ & $0.251 _{-0.007}^{+0.007}$ & $16.20^{-1.80}_{1.72}$\\ 
  FW Cas & 6.2400 & 6.1099 & 1.50 & -0.12 &  0.02 & 0.28 & $5.0_{-0.5}^{+0.5}$ & $5500 _{-50}^{+50}$ & $3.26 _{-0.04}^{+0.04}$ & $0.19 _{-0.10}^{+0.10}$ & $1.15_{-0.12}^{+0.13}$ & $-50 _{-1}^{+1}$ & $0.260 _{-0.009}^{+0.009}$ & $13.53^{-1.08}_{0.99}$\\ 
  AT Pup & 6.6656 & 6.5634 & 1.40 & -0.08 & 0.02 & 0.28 & $5.5 _{-0.5}^{+0.5}$ & $5500 _{-50}^{+50}$ & $3.33 _{-0.04}^{+0.04}$ & $0.12 _{-0.12}^{+0.12}$ & $1.17_{-0.19}^{+0.22}$ & $25 _{-3}^{+4}$ & $0.655 _{-0.009}^{+0.009}$ & $90.30^{-50.38}_{43.64}$\\ 
  V737 Cen & 7.0658 & 7.0748 & 1.55 & 0.11 & 0.03 & 0.28 & $6.5 _{-0.5}^{+0.5}$ & $5250 _{-50}^{+50}$ & $3.34 _{-0.02}^{+0.02}$ & $-0.048 _{-0.015}^{+0.015}$ & $1.16_{-0.08}^{+0.08}$ & $8.2 _{-0.8}^{+0.7}$ & $1.285 _{-0.032}^{+0.032}$ & $7.61^{-2.49}_{2.70}$\\ 
  ASASJ084412 & 7.2154 & 7.1754 & 1.60 & 0.12 & 0.03 & 0.28 & $6.0 _{-0.5}^{+0.5}$ & $5250 _{-50}^{+50}$ & $3.32 _{-0.01}^{+0.01}$ & $0.049 _{-0.007}^{+0.007}$ & $1.20_{-0.29}^{+0.23}$ & $58.6 _{-0.8}^{+0.9}$ & $0.332 _{-0.011}^{+0.011}$ & $6.86^{-0.47}_{0.42}$\\ 
  MN Cam & 8.1796 & 7.8565 & 1.50 & -0.02 &  0.01 & 0.26 & $6.5 _{-0.5}^{+0.5}$ & $5450 _{-50}^{+50}$ & $3.47 _{-0.01}^{+0.01}$ & $-0.039 _{-0.007}^{+0.007}$ & $1.16 _{-0.16}^{+0.18}$ & $-58 _{-1}^{+1}$ & $0.205 _{-0.014}^{+0.014}$ & $13.90^{-2.91}_{3.25}$\\ 
  WX Pup & 8.9364 & 8.4958 & 1.45 & -0.01 & 0.01 & 0.26 & $6.0_{-0.5}^{+0.5}$ & $5550 _{-100}^{+100}$ & $3.51 _{-0.02}^{+0.02}$ & $0.12 _{-0.07}^{+0.07}$ & $1.17 _{-0.16}^{+0.13}$ & $54 _{-1}^{+1}$ & $0.400 _{-0.003}^{+0.003}$ & $13.34^{-7.17}_{10.25}$\\ 
  CF Cam & 9.4363 & 8.9451 & 1.70 & -0.17 & 0.01 & 0.26 & $7.0_{-0.5}^{+0.5}$ & $5500 _{-75}^{+75}$ & $3.57 _{-0.02}^{+0.02}$ & $-0.04 _{-0.02}^{+0.02}$ & $1.30 _{-0.27}^{+0.30}$ & $-36 _{-1}^{+1}$ & $0.274 _{-0.011}^{+0.011}$ & $2.22^{-0.65}_{0.78}$\\ 
  V339 Cen & 9.4662 & 9.2951 & 1.60 & 0.06 & 0.02 & 0.28 & $6.5_{-0.5}^{+0.5}$ & $5200 _{-50}^{+50}$ & $3.46 _{-0.02}^{+0.02}$ & $0.015 _{-0.015}^{+0.015}$ & $1.49 _{-0.41}^{+0.40}$ & $-24 _{-2}^{+2}$ & $0.614 _{-0.014}^{+0.014}$ &$88.48^{-54.85}_{56.83}$ \\ 
  ASASJ171305 & 9.5614 & 9.2303 & 1.40 & 0.35  & 0.01 & 0.26 & $6.5 _{-0.5}^{+0.5}$ & $5300 _{-50}^{+50}$ & $3.5 _{-0.02}^{+0.02}$ & $-0.006 _{-0.014}^{+0.014}$ & $1.40_{-0.23}^{+0.24}$ & $-19 _{-1}^{+1}$ & $0.480 _{-0.039}^{+0.039}$ & $5.37^{-4.91}_{4.34}$\\ 
  OGLE0339 & 10.1737 & 10.2435 & 1.50 & 0.29 & 0.03 & 0.28 & $7.0 _{-0.5}^{+0.5}$ & $5000 _{-50}^{+50}$ & $3.47 _{-0.03}^{+0.03}$ & $-0.03 _{-0.08}^{+0.08}$ & $1.32 _{-0.56}^{+0.57}$ & $31 _{-6}^{+5}$ & $1.233 _{-0.256}^{+0.256}$ & $34.97^{-10.00}_{10.79}$\\ 
  HZ Per & 11.2791 & 11.0113 & 1.80 & -0.28 & 0.01 & 0.26 & $5.5 _{-0.5}^{+0.5}$ & $5350 _{-50}^{+50}$ & $3.54 _{-0.03}^{+0.03}$ & $0.28 _{-0.11}^{+0.11}$ & $1.16 _{-0.28}^{+0.33}$ & $-23 _{-6}^{+9}$ & $0.228 _{-0.014}^{+0.014}$ & $222.70^{-2.39}_{3.46}$\\ 
  SS CMa & 12.3523 & 12.1861 & 1.60 & 0.06 & 0.02 & 0.28 & $5.0 _{-1}^{+1}$ & $4950 _{-250}^{+250}$ & $3.46 _{-0.12}^{+0.12}$ & $0.25 _{-0.27}^{+0.27}$ & $1.17 _{-0.49}^{+0.37}$ & $77 _{-8}^{+6}$ & $0.432 _{-0.018}^{+0.018}$ & $42.93^{-84.03}_{73.59}$\\ 
  AD Pup & 13.5974 & 13.4000 & 1.60 & -0.06 & 0.02 & 0.28 & $7.0 _{-0.5}^{+0.5}$ & $4950 _{-50}^{+50}$ & $3.59 _{-0.03}^{+0.03}$ & $0.03 _{-0.10}^{+0.10}$ & $1.20 _{-0.037}^{+0.038}$ & $78 _{-1}^{+1}$ & $0.284 _{-0.012}^{+0.012}$ & $45.69^{-1.67}_{1.54}$\\ 
  TX Cyg & 14.7122 & 15.1100 & 1.50 & 0.26 & 0.04 & 0.29 & $8.0 _{-0.5}^{+0.5}$ & $4800 _{-50}^{+50}$ & $3.63 _{-0.01}^{+0.01}$ & $-0.04 _{-0.04}^{+0.04}$ & $1.17_{-0.06}^{+0.07}$ & $-17 _{-1}^{+1}$ & $1.161 _{-0.082}^{+0.082}$ & $8.29^{-5.05}_{5.80}$\\ 
  RW Cas & 14.7889 & 15.2405 & 1.45 & 0.28 & 0.04 & 0.29 & $7.0 _{-0.75}^{+0.75}$ & $4700 _{-50}^{+50}$ & $3.58 _{-0.04}^{+0.04}$ & $0.08 _{-0.12}^{+0.12}$ & $1.16 _{-0.15}^{+0.22}$ & $-68 _{-2}^{+3}$ & $0.463 _{-0.027}^{+0.027}$ & $11.00^{-59.84}_{39.54}$\\ 
  RZ Vel & 20.4289 & 20.5388 & 1.50 & 0.19 & 0.03 & 0.28 & $9.0 _{-0.5}^{+0.5}$ & $4800 _{-50}^{+50}$ & $3.83 _{-0.02}^{+0.02}$ & $-0.04 _{-0.06}^{+0.06}$ & $1.15 _{-0.07}^{+0.07}$ & $27 _{-3}^{+4}$ & $0.765 _{-0.031}^{+0.031}$ & $21.47^{-11.62}_{14.44}$\\ 
  S Vul & 69.4674 & 71.2754 & 1.50 & 0.09 & 0.02 & 0.28 & $11.0 _{-0.5}^{+0.5}$ & $4300 _{-50}^{+50}$ & $4.31 _{-0.02}^{+0.02}$ & $0.10 _{-0.03}^{+0.03}$ & $1.17 _{-0.18}^{+0.18}$ & $4 _{-1}^{+1}$ & $0.408 _{-0.023}^{+0.023}$ & $21.52^{-2.22}_{2.26}$\\ 
   \hline
\end{tabular}
\tablefoot{Column 1 lists the names of the sources. Columns 2 and 3 provide the pulsational periods from observations and the best-modeled periods, respectively. Column 4 contains the best-fit $\alpha$ values. Column 5 presents the spectroscopic $\rm [FeH]$ values from T24. Columns 6 and 7 report the best-modeled elemental compositions, $\rm Z_{best}$ and $\rm Y_{best}$, respectively. Columns 8–11 include the best-fit stellar masses, effective temperatures, luminosities, and non-canonical over-luminosities. Columns 12 and 13 list the best-fit projection factor values and barycentric $\gamma$ velocities. Finally, column 14 provides the best estimates of the source parallaxes, while column 15 list the $\chi^2$ minimum values with their errors, obtained by using our fitting method.}
\label{tab:bestFitParams}
\end{table*}
\end{appendix}

\end{document}